\documentclass[letterpaper, 10pt]{article}
\usepackage{graphicx}
\usepackage{amsmath}
\usepackage{amsfonts}
\usepackage{hyperref}
\usepackage{amsthm}
 \usepackage{amscd}
\usepackage{mathrsfs}
\usepackage{caption}
\usepackage{subcaption}
\usepackage{float}
\usepackage[top=1in,left=1in,right=1in,bottom=1in]{geometry}

\hypersetup{
	colorlinks = true,
	linkcolor  = red,
	citecolor  = blue,
	filecolor  = cyan,
	urlcolor   = magenta,}

\newcommand{\ds}{\displaystyle}

\numberwithin{equation}{section}

\title{The Marchenko method for soliton solutions to the Sawada--Kotera equation\thanks{Dedicated to the memory of David J. Kaup, who started the study of the
inverse scattering problem for the third-order equation}}
\author{Tuncay Aktosun\\
Department of Mathematics\\
University of Texas at Arlington\\
Arlington, TX 76019-0408, USA\\
\\
Ramazan Ercan\\
Department of Mathematics\\
California State University San Marcos\\
San Marcos, CA 92096, USA\\
\\
Ivan Toledo\\
Department of Mathematics\\
University of Texas at Arlington\\
Arlington, TX 76019-0408, USA\\
\\
Mehmet Unlu\thanks{Corresponding author: mehmet.unlu@erdogan.edu.tr}\\
Department of Mathematics\\
Recep Tayyip Erdogan University\\
53100 Rize, Turkey}

\date{}

\begin{document}

\maketitle

\begin{abstract}
Associated with the third-order linear differential operator,
we present the Marchenko integral equation using as input the bound-state
poles of a transmission coefficient and the time-evolved bound-state dependency
constants. We derive 
the $\mathbf N$-soliton solution to the Sawada--Kotera equation, for an arbitrary positive
integer $\mathbf N,$ by recovering that soliton solution from the solution to our Marchenko integral equation.
Our method explains the origin of the $2\mathbf N$ real parameters appearing 
in the $\mathbf N$-soliton solution formula obtained by the ad-hoc method of Hirota.
We show that $\mathbf N$ of those parameters are related to the $\mathbf N$ bound-state poles of
the left transmission coefficient and the remaining $\mathbf N$ 
parameters are related to the bound-state dependency
constants. Our Marchenko integral equation corresponds to the ``GLM (Gel'fand--Levitan--Marchenko) integral equation'' Kaup
relentlessly but unsuccessfully tried to obtain.

\end{abstract}

{\bf {AMS Subject Classification (2020):}} 34A55 34M50 35C08 37K10

{{\bf Keywords:} Sawada--Kotera equation, soliton solutions, Marchenko method, Marchenko integral equation, inverse scattering for the third-order operator}

\newpage

\section{Introduction}
\label{section1} 

In this paper we present the derivation of the $\mathbf N$-soliton solution, where
$\mathbf N$ is an arbitrary positive integer,
to the integrable evolution equation known as the Sawada--Kotera equation \cite{SK1974}, i.e.
\begin{equation}\label{1.1}
Q_t+Q_{xxxxx}+5\,Q_x\,Q_{xx}+5\,Q\,Q_{xxx}+5\,Q^2\,Q_x=0, \qquad x,\,t\in\mathbb R,
\end{equation}
with $x$ and $t$ being the spacial and temporal independent variables, respectively, taking values on the real line $\mathbb R.$ The subscripts
in \eqref{1.1} denote the corresponding partial derivatives.
The mathematical significance of our paper comes from the fact that we obtain the $\mathbf N$-soliton solution to
\eqref{1.1} from the solution to a linear integral equation
associated with the inverse
scattering problem for the third-order linear differential operator on the full line.
We refer to our linear integral equation as the Marchenko integral equation.

A Marchenko integral equation \cite{AK2001,CS1989,DT1979,L1987,M1955,M2011,N1983} is used to solve the inverse scattering
problems related to various linear differential operators such as those associated with the KdV (Korteweg--de Vries)
equation  \cite{KdV1895} 
and the NLS (nonlinear Schr\"odinger)
equation \cite{ZS1972}. The Marchenko integral equation for the KdV equation
arises in the analysis of the inverse scattering problem for
the second-order linear differential operator, and
the Marchenko system of linear integral equations arises in
the analysis of a linear differential operator associated with a system of first-order equations.
Our primary goal in this paper is to establish the Marchenko integral equation corresponding to the $\mathbf N$-soliton solution to \eqref{1.1} in such a way that our method also applies to other integrable evolution equations associated with the
third-order linear differential operator, and those evolution equations
include the Kaup–Kupershmidt equation \cite{K1980,K2002,K1984}, the good Boussinesq equation \cite{CL2022}, the bad Boussinesq
equation \cite{B1872,CL2024}, and a modified version of the bad Boussinesq equation not containing the second $x$-derivative term \cite{DTT1982,M1978}.

In order to obtain solutions to \eqref{1.1}, Kaup initiated \cite{K1980} the study of the inverse scattering problem for the third-order linear
operator. He derived \cite{K1980} the one-soliton solution to the Sawada--Kotera equation as well as the one-soliton solution to the
Kaup--Kupershmidt equation. He tried to formulate the scattering theory
for the third-order operator, but as he indicated \cite{K1980} he was unsuccessful even in defining the scattering coefficients properly. The ``scattering matrix'' $a_{mn}$ introduced in (2.9)--(2.11) of \cite{K1980} is not the analog of the
scattering matrix $S(k)$ used in the second-order case, but it is the analog of
the transition matrix $\Lambda(k)$ used in the second-order case. For the distinction
between the scattering matrix $S(k)$ and the transition matrix $\Lambda(k)$  in the second-order case,
we refer the reader to (1.2) of \cite{A1992} for $S(k)$ and to (2.1) of \cite{A1992}
 for $\Lambda(k).$ On p.~190 in \cite{K1980}, Kaup stated that
``Due to its complexity, at the present time we leave many
questions unanswered. We shall point out what these are in a summary at the
end.'' Kaup tried to deal with various other difficulties. On p.~203 in \cite{K1980}, Kaup stated that 
``With all of this, we may now obtain integral `dispersion relations' for all of our
eigenfunctions in terms of the others and the scattering data. To do this, we shall
assume compact support, so that all functions are entire.''
For other difficulties Kaup encountered, we refer the reader to Sections~III and V of his 1980
paper \cite{K1980}. On p.~2706 of \cite{H1989}, Hirota stated that
``Kaup \cite{K1980} tried to solve the inverse scattering problem of the class
$\psi_{xxx}+6Q\psi_x+6R\psi=\lambda\psi,$ but was unsuccessful.''

Kaup relentlessly but unsuccessfully tried \cite{K2002} to find the analog of the Marchenko integral equation corresponding to \eqref{1.1}. Kaup's strong desire
and unsuccessful efforts in this direction were described in the last section entitled ``Unsolved Problems'' in his 2002 paper
\cite{K2002}. He stated that ``Now we will discuss what the hard problems are. They are hard because they haven’t been solved. This is
not a complete list, but is the start of a list of problems in need of solutions. Although we have a solution for the IST of the cubic eigenvalue
problem, there still is no equivalent of the GLM equations for this system. What we do have is the solution of the Riemann--Hilbert problem
for the eigenfunctions. What is next needed is a representation of the eigenfunctions in terms of some transformation kernels
(like the $K(x, y; t)$ for the KdV). Once these are known, then the equivalent GLM equations will follow.'' The GLM
(Gel'fand--Levitan--Marchenko) equation Kaup wanted to obtain is exactly the Marchenko equation we present in our paper for the
Sawada--Kotera equation \eqref{1.1} with its $\mathbf N$-soliton solutions. The IST of the cubic eigenvalue problem mentioned by
Kaup refers to the inverse scattering transform for the third-order linear differential operator.

A Marchenko integral equation is at times referred to as the GLM equation, as Kaup mentioned it in \cite{K2002}, but this is a misnomer \cite{N1980}. The
Gel'fand--Levitan integral equation and the Marchenko integral equation are two separate integral equations \cite{CS1989,L1987,M2011,N1983} used to solve
various inverse spectral and scattering problems, respectively. The kernel and the nonhomogeneous term in the Gel'fand--Levitan integral equation \cite{GL1955} are related to the
spectral measure \cite{AK2001,CS1989,GL1955,L1987,M2011,N1983}. The Gel'fand--Levitan integral equation is normally used to solve an inverse problem on a half line or on a finite interval, and the
integral in the Gel'fand--Levitan integral equation is over a finite interval such as $[0,x].$ On the other hand, the kernel and the
nonhomogeneous term in the Marchenko integral equation \cite{DT1979,F1967,M1955} are related to the Fourier transform of the scattering matrix or of a reflection coefficient and the
bound-state information 
\cite{AK2001,CS1989,GL1955,L1987,M2011,N1983}. The Marchenko integral equation is used to solve an inverse scattering problem either on a half line or on the full
line, and the integral in the Marchenko equation is over a semi-infinite interval containing either $+\infty$ or $-\infty$ at one of its end points.

The $\mathbf N$-soliton solution to the Sawada--Kotera equation was first provided by Hirota \cite{H1989}, in which the dressing method \cite{ZS1974} of
Shabat and Zakharov was used. In \cite{H1989}, Hirota formulated an integral equation whose solution yields soliton solutions to the
BKP (B-type Kadomtsev--Petviashvili) equation. In that $\mathbf N$-soliton solution \cite{H1989}, by letting $y=0$ and using a partial integration
in the $x$-variable, one obtains the $\mathbf N$-soliton solution to the Sawada--Kotera equation \eqref{1.1}. However, Hirota's
integral equation
presented in \cite{H1989} is not the analog of a Marchenko integral equation for which Kaup was looking.
There are two main issues with Hirota's integral equation. The first issue is that its integrand involves Hirota's bilinear form \cite{H2004}. The second
issue is that the kernel and the nonhomogeneous term in that integral equation are specified in an ad-hoc manner with no connection to the
third-order linear differential operator or to any physical quantities. In other words, the input to Hirota's integral equation does not use any scattering data set related to
the third-order operator. Unsatisfied with the complications due to the use of Hirota's bilinear form, Hereman and Nuseir
\cite{HN1997} provided a simplified version of Hirota's method without needing a bilinear representation. Although the $\mathbf N$-soliton solution
is formulated by the simplified version of Hirota's bilinear method, it still remains an ad-hoc method, and the parameters appearing in the
$\mathbf N$-soliton solution have not been shown to be related to any physical parameters.
The unsatisfactory aspects of Hirota's integral equation in \cite{H1989} motivated Parker \cite{P2001} to try to use a modification of the dressing method
so that Hirota's integral equation could be transformed into a linear integral equation resembling a Marchenko integral equation.
Although Parker was not successful, his paper \cite{P2001} is still informative and relevant. 
 
We refer the reader to \cite{ACTU2025,C1980,DTT1982,K1980,T2024} for the general study where the scattering and inverse scattering problems associated with the third-order linear operator are analyzed.
We recently presented a method \cite{ACTU2025,ACTU2026} for the derivation of the $\mathbf N$-soliton solution formula for the Sawada--Kotera equation by analyzing the inverse scattering
problem for the third-order linear differential operator. In this method,
by formulating a Riemann--Hilbert problem in the complex plane and by solving that Riemann--Hilbert problem in a closed form, we
obtain an explicit expression for the $\mathbf N$-soliton solution, for any positive integer $\mathbf N,$ to the Sawada--Kotera equation. The input used to solve the aforementioned
Riemann--Hilbert problem consists of the bound-state information for the third-order operator
in the reflectionless case, namely the bound-state values 
of the spectral parameter and the bound-state dependency constants. A summary of the derivation
of that $\mathbf N$-soliton solution formula
via our formulation of the 
Riemann--Hilbert problem is presented in Section~\ref{section4} in our present paper. We remark that, in solving 
inverse scattering problems, it is customary \cite{CS1989,L1987,M2011} to formulate a Riemann--Hilbert problem by recovering a sectionally
analytic function from the jump determined by the scattering coefficients and the bound-state information.
In our own formulation of the Riemann--Hilbert problem for the third-order linear operator, we recover
a sectionally meromorphic function from the jump solely determined by the reflection coefficients.
This approach, i.e. the detachment of the bound-state information from the jump,
offers a significant advantage in solving the inverse scattering problem for the third-order linear operator
in the reflectionless case.

In the present paper, by using the appropriate Fourier transformation on our aforementioned
 Riemann--Hilbert problem formulated in \cite{ACTU2025,ACTU2026}, we derive
our Marchenko integral equation associated with the $\mathbf N$-soliton solution to the Sawada--Kotera equation. The input to our Marchenko integral equation is the same input used for our Riemann--Hilbert problem. In fact, we show that our relevant input data set
can be expressed in terms of $2\mathbf N$ positive parameters, as in the case where a typical $\mathbf N$-soliton solution formula to an
integrable evolution equation is expressed in a closed form by using $2\mathbf N$ parameters
associated with $\mathbf N$ simple bound-state energy values and $\mathbf N$ time-evolved bound-state normalization constants.
We refer to this procedure as the Marchenko method as it is similar to the procedure used in the Marchenko method to solve other
integrable evolution equations \cite{AC1991,AS1981,A2009,L1980,NMPZ1984}
via the inverse scattering transform method \cite{GGKM1967}.
In this case, our Marchenko integral equation has a separable kernel, and hence it can be solved explicitly in a closed form by using the
methods of linear algebra. We then show how the $\mathbf N$-soliton solution formula is obtained from the solution to our Marchenko integral equation. By comparing the $\mathbf N$-soliton solution formula obtained from the solution
to our Riemann--Hilbert problem or the solution to our Marchenko integral equation
with the corresponding expression obtained from Hirota's method,
we reveal how the parameters in Hirota's method are related to the bound-state information for the third-order 
linear operator
in the reflectionless case.

Our paper is organized as follows. In Section~\ref{section2} we provide a summary of the connection between
the Sawada--Kotera equation and the third-order linear differential operator, by making the distinction between
the two cases SK1 and SK2, where the two potentials in the linear equation are related to each other in two specific
manners. In Section~\ref{section3} we briefly describe the scattering problem
for the third-order linear operator in the reflectionless case, and we
indicate how the two potentials are obtained from the large spectral asymptotics 
of the so-called left and right Jost solutions to the third-order linear equation.
In Section~\ref{section4}, we establish our Riemann--Hilbert problem
associated with the inverse scattering problem for the third-order linear operator
in the reflectionless case. We use as input the input data set consisting of the bound-state poles and
the bound-state dependency constants, and we show how the solution to the Riemann--Hilbert problem
yields the two potentials in the third-order linear operator. In Section~\ref{section5}, we show how the restrictions
on the time-evolved dependency constants yield the $\mathbf N$-soliton solution to the
Sawada--Kotera equation.
In Section~\ref{section6}, we derive our
Marchenko integral equation associated with
the third-order linear operator in the reflectionless case. This is done by using
the appropriate Fourier transformation on
our Riemann--Hilbert problem. By using the same input used in the Riemann--Hilbert problem,
we show how the solution to the Marchenko integral equation yields
the potentials in the third-order operator in the reflectionless case.
We then show that the appropriate restrictions on the time-evolved dependency
constants yield
the $\mathbf N$-soliton solution to the
Sawada--Kotera equation.
 Finally, in Section~\ref{section7} we provide the conclusion by
summarizing the significance of our results and the relevant planned work in the near future.

To help the reader quickly concentrate on the main issue in our paper, i.e. the establishment of 
the Marchenko integral equation and the construction of the $\mathbf N$-soliton solution to the Sawada--Kotera equation,
we provide the following guidance.
We associate the Sawada--Kotera equation \eqref{1.1} with two cases of the third-order linear equation
\eqref{2.4}. We refer to the first case as the SK1 case and to the second as the SK2 case.
In both cases, the potential $Q$ in \eqref{2.4} corresponds to the unknown $Q$ in the Sawada--Kotera equation.
In the SK1 case the potential $P$ in \eqref{2.4} is given by $P\equiv 0$
and in the SK2 case the potential $P$ in \eqref{2.4} is related to $Q$ as $P = Q_x.$
In Section~\ref{section6} we establish the Marchenko integral equation \eqref{6.29} associated with the third-order linear 
equation \eqref{2.4}. The kernel $\Omega(\zeta,y)$ in that Marchenko integral equation is given in \eqref{6.30}
and that kernel is constructed by using the bound-state poles of the left transmission coefficient 
 $T_{\text{\rm{l}}}(k)$ and the time-evolved
bound-state dependency constants $D_j$ and $D_j^\ast$
in \eqref{4.6} and \eqref{4.7},
respectively, for $1\le j\le \mathbf N.$
The nonhomogeneous term in the Marchenko integral
equation is given by $\Omega(0,y)$
and hence that nonhomogeneous term is also constructed by using the same input data set 
used to construct
the integral kernel.
The time-evolved dependency constants $D_j$ and $D_j^\ast$
can equivalently be expressed in terms of the time-independent dependency constants 
$E_j$ and $E_j^\ast$ as in \eqref{4.8}
or they can be expressed in terms of the modified-dependency constants $\gamma_j$
and $\gamma_j^\ast$ appearing in \eqref{4.20} and \eqref{5.1}.
We use an asterisk to denote complex conjugation.
As in \eqref{5.1}, we write $\gamma_j$ in terms of its real and imaginary parts
$r_j$ and $s_j,$ respectively.
The modified dependency constants $\gamma_j$ and $\gamma_j^\ast$
are uniquely determined in each of the SK1 and SK2 cases by using the ratios $s_j/r_j$
in \eqref{5.34} and \eqref{5.35}, respectively, in terms of the real parameters $r_j$
for $1\le j\le \mathbf N.$
In fact, instead of using the real parameters $r_j,$ we can use the positive parameters $c_j$
given in \eqref{5.42} and \eqref{5.43} in the SK1 and SK2 cases, respectively.
Thus, the input to the Marchenko integral equation is equivalent to
$\{k_j,c_j\}_{j=1}^{\mathbf N}.$ Consequently, in each of the cases of SK1 and SK2 we have the 
equivalence of the input data sets indicated in the diagram
\begin{equation}
\label{1.2}
\{k_j,D_j\}_{j=1}^{\mathbf N}\longleftrightarrow\{k_j,E_j\}_{j=1}^{\mathbf N}\longleftrightarrow
\{k_j,\gamma_j\}_{j=1}^{\mathbf N}\longleftrightarrow\{k_j,r_j\}_{j=1}^{\mathbf N}\longleftrightarrow\{k_j,c_j\}_{j=1}^{\mathbf N}.
\end{equation} 
Using our input data set, we construct the kernel 
$\Omega(\zeta,y)$ 
and the nonhomogeneous term
$\Omega(0,y)$ 
of the Marchenko integral equation \eqref{6.29}.
In this case, 
the Marchenko integral equation has a separable kernel and hence
we obtain the solution to \eqref{6.29} explicitly in a closed form, by using the methods of linear algebra, as in \eqref{6.44}.
We then recover the potential $Q$ as in \eqref{6.48}, 
where 
$\det[\mathbf m(x)]$ is the quantity
expressed as in \eqref{6.50} and
$\Delta(x)$ 
is the quantity in \eqref{5.44}, explicitly expressed in terms of the input data set 
$\{k_j,c_j\}_{j=1}^{\mathbf N}.$
Thus,
the potential $Q,$ constructed as in \eqref{6.48} or equivalently as in \eqref{5.46}, yields the 
$\mathbf N$-soliton solution to the Sawada--Kotera equation \eqref{1.1}.

\section{The Sawada--Kotera equation}
\label{section2} 

In this section we relate the Sawada--Kotera equation \eqref{1.1} to the third-order ordinary differential operator
associated with \eqref{2.4}. This
is because our goal is to obtain soliton solutions to \eqref{1.1} by analyzing the scattering and inverse scattering for the corresponding third-order
linear operator in the reflectionless case. For the analysis of the scattering problem for the third-order operator when
the reflection coefficients are not zero, we refer the reader to \cite{ACTU2025,ATU2025}.

Being an integrable nonlinear evolution equation, the Sawada--Kotera equation \eqref{1.1} can be derived \cite{ACTU2025} by using either of the two Lax pairs $(L_1,A_1)$ and $(L_2,A_2)$ given by
 \begin{equation}
 \label{2.1}
 \begin{cases}
L_1:=D^3+QD,\\
\noalign{\medskip}
A_1:=9\,D^5+15\,Q D^3+15\,Q_x\,D^2+\big(10\,Q_{xx}+5\,Q^2\big)D,
 \end{cases}     
\end{equation}
 \begin{equation}
 \label{2.2}
 \begin{cases}
L_2:=D^3+QD+Q_x,\\
\noalign{\medskip}
A_2:=9\,D^5+15\,Q D^3+30\,Q_x\,D^2+\big(25\,Q_{xx}+5\,Q^2\big)D
      +\big(10\,Q_{xxx}+10\,Q \,Q_x\big),
 \end{cases}     
\end{equation}
where we use $D$ to denote the derivative operator by letting $D:=d/dx$ and $D^n:=d^n/dx^n$ for $n\ge 2.$ 
In other words, the two Lax operator equations \cite{L1968}
 \begin{equation}
  \label{2.3}
\dot L_1+L_1 A_1-A_1 L_1=0,
\quad
\dot L_2+L_2 A_2-A_2 L_2=0,
\end{equation}
where the overdot denotes the $t$-derivative, are satisfied provided that the Sawada--Kotera equation \eqref{1.1}
is satisfied.
Letting $L\psi=k^3\psi,$ we see that
the Sawada--Kotera
equation \eqref{1.1} is associated with the third-order ordinary linear differential equation
\begin{equation}
\label{2.4}
\psi'''+Q\,\psi'+P\,\psi=k^3\,\psi, \qquad x\in\mathbb R,
\end{equation}
where $\psi$ is the wavefunction, $k^3$ is the spectral parameter, $x$ is the independent variable, the prime denotes the $x$-derivative, $t$ now appears as a parameter, and the coefficients
$Q$ and $P$ are real-valued functions of the independent variable $x$ and the parameter $t.$ We refer to the coefficients
$Q$ and $P$ as the potentials. The dependence of $Q$ and $P$ on the parameter $t$ is governed by either of the linear operators $A_1$ and $A_2$
in \eqref{2.1} and \eqref{2.2}, respectively. For simplicity, we assume that $Q$ and $P$ in \eqref{2.4} belong to the Schwartz class in $x\in\mathbb R$ for each fixed $t$ even though our results hold under weaker restrictions on $Q$ and $P.$

The direct scattering problem for \eqref{2.4} consists of the determination of the scattering coefficients and the bound-state information when the
potentials $Q$ and $P$ are known. On the other hand, the inverse scattering problem for \eqref{2.4} involves the recovery of $Q$ and $P$ from
an appropriate input data set including the scattering coefficients and the bound-state information.

From \eqref{2.1}, \eqref{2.2}, and \eqref{2.4} we see that the linear operator $L_1$
corresponds to the case $P\equiv 0$ and the linear operator $L_2$ corresponds to the case $P=Q_x$ and that the same $Q$ appears
in both $L_1$ and $L_2.$ That same $Q$ also appears in the Sawada--Kotera equation \eqref{1.1} whereas $P$ does not appear in
\eqref{1.1} at all. Consequently, if \eqref{1.1} is analyzed without connecting it to the inverse scattering problem for \eqref{2.4}, we do not need to
make a distinction between the case $P\equiv 0$ and the case $P=Q_x.$ On the other hand, if the solution to \eqref{1.1} involves a method
related to
the solution of the inverse scattering problem for \eqref{2.4}, we need to make a distinction between those two cases. 
We refer the reader to \cite{ACTU2025,ACTU2026} for more information on this issue.
As already mentioned, we use SK1 to refer to the case $P\equiv 0$ and we use SK2 to refer to the case $P=Q_x.$

For the analysis of soliton solutions to the Sawada--Kotera equation \eqref{1.1},
the aforementioned distinction is as follows.
When $Q$ is real valued, it is known \cite{ACTU2025,ACTU2026} that the third-order equation \eqref{2.4} in the cases of $P\equiv 0$ and $P=Q_x,$ respectively, comprises two
linear third-order adjoint equations, for which the corresponding transmission coefficients coincide and hence also the bound-state spectral values for the two equations coincide. On the other hand, the bound-state dependency constant at each bound state is not the same for the two equations. 
Hence, both of the cases $P\equiv 0$ and $P=Q_x$ in \eqref{2.4} can be analyzed by using
the zero reflection coefficients, the same set of  left and right transmission coefficients, but two different sets of bound-state dependency constants. In our paper, we derive the $\mathbf N$-soliton solution to
\eqref{1.1} by solving the inverse scattering problem for \eqref{2.4} in the reflectionless case. We use the input data set consisting of a particular left transmission
coefficient, the bound-state poles of that left transmission coefficient, and the dependency constants at those bound-state poles
by explicitly mentioning whether we use the bound-state dependency constants for SK1 or for SK2.

\section{The third-order equation in the reflectionless case}
\label{section3}

In this section we present a summary of the basic results for the direct scattering problem for \eqref{2.4} in the reflectionless case. 
We refer the reader to \cite{ACTU2025,ATU2025} for the general case when the reflection coefficients are not zero.
We recall that we assume that the potentials $Q$ and $P$ in
\eqref{2.4} belong to the Schwartz class in $x\in\mathbb R$ for each fixed $t.$

\begin{figure}[!ht]
     \centering
         \includegraphics[width=2.in,height=3.4in]{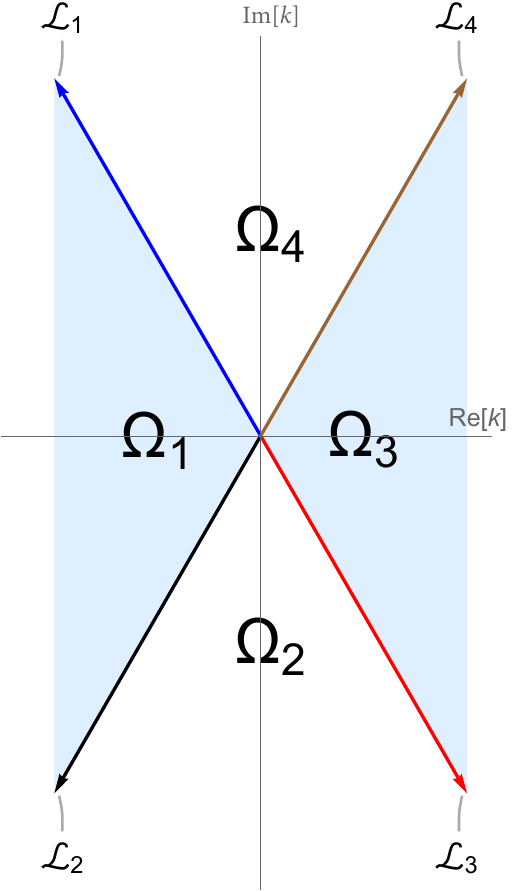}      \hskip .2in
         \includegraphics[width=2.in,height=3.4in]{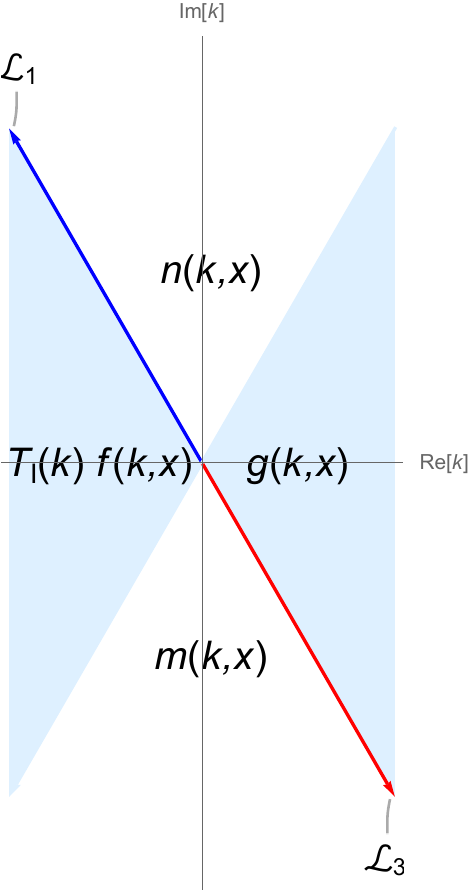}      \hskip .2in
         \includegraphics[width=2.in,height=3.4in]{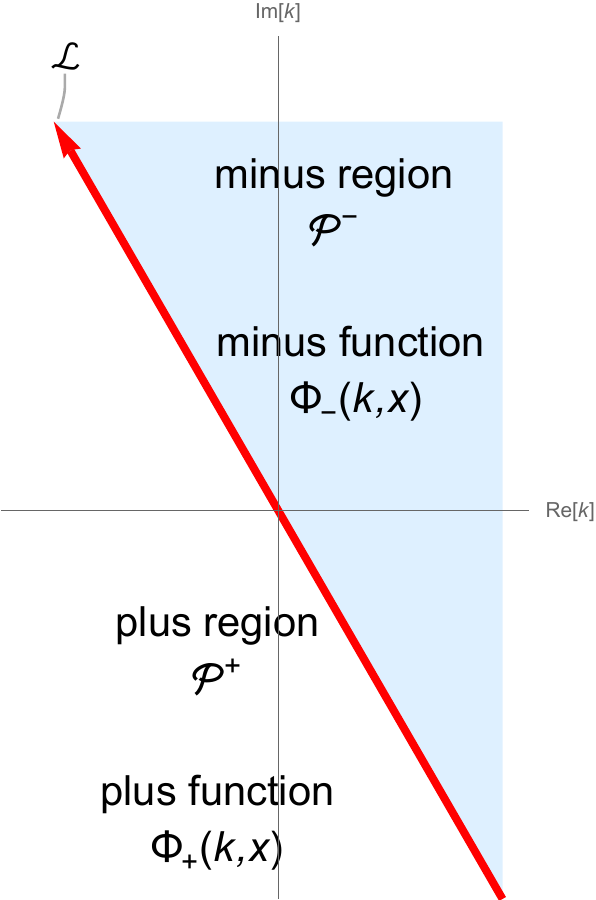} 
\caption{The complex $k$-plane is divided into the 
four sectors
 $\Omega_1,$ 
$\Omega_2,$ $\Omega_3,$ and $\Omega_4$ as shown on the left plot,
with the directed half lines $\mathcal L_1,$ $\mathcal L_2,$ $\mathcal L_3,$ and $\mathcal L_4$
acting as the boundaries. The $k$-domains of $T_{\text{\rm{l}}}(k) f(k,x),$ $m(k,x),$ $g(k,x),$ and $n(k,x),$ respectively, are shown on the middle
plot. The right plot shows the plus and minus regions in the complex $k$-plane separated by the directed full line $\mathcal L,$
as well as the plus and minus functions in their respective $k$-domains.}
\label{Figure1}
\end{figure}

We divide the complex $k$-plane into four open sectors $\Omega_1,$ $\Omega_2,$ $\Omega_3,$ $\Omega_4$ as indicated
on the left plot of Figure~\ref{Figure1} by using the directed half lines $\mathcal L_1,$ $\mathcal L_2,$ $\mathcal L_3,$ $\mathcal L_4,$ which are
parametrized as
\begin{equation}\label{3.1}
\mathcal L_1:=\{k\in\mathbb C: k=zs \text{\rm{ for }}  s\in[0,+\infty)\},
\end{equation}
\begin{equation*}
\mathcal L_2:=\{k\in\mathbb C: k=z^2s \text{\rm{ for }}  s\in[0,+\infty)\},
\end{equation*}
\begin{equation}\label{3.3}
\mathcal L_3:=\{k\in\mathbb C: k=-zs \text{\rm{ for }}  s\in[0,+\infty)\},
\end{equation}
\begin{equation*}
\mathcal L_4:=\{k\in\mathbb C: k=-z^2s \text{\rm{ for }}  s\in[0,+\infty)\},
\end{equation*}
where $z$ is used to denote the special complex number $e^{2\pi i/3},$ which is also expressed as
\begin{equation}\label{3.5}
z:=-\displaystyle\frac{1}{2}+i\,\displaystyle\frac{\sqrt{3}}{2}.
\end{equation}
The open sectors $\Omega_1,$ $\Omega_2,$ $\Omega_3,$ $\Omega_4$ are described by using
the parametrizations given by
\begin{equation*}
\Omega_1:=\left\{k\in\mathbb C: \displaystyle\frac{2\pi}{3}<\arg[k]<\displaystyle\frac{4\pi}{3}\right\},
\end{equation*}
\begin{equation*}
\Omega_2:=\left\{k\in\mathbb C: -\displaystyle\frac{2\pi}{3}<\arg[k]<-\displaystyle\frac{\pi}{3}\right\},
\end{equation*}
\begin{equation*}
\Omega_3:=\left\{k\in\mathbb C: -\displaystyle\frac{\pi}{3}<\arg[k]<\displaystyle\frac{\pi}{3}\right\},
\end{equation*}
\begin{equation*}
\Omega_4:=\left\{k\in\mathbb C: \displaystyle\frac{\pi}{3}<\arg[k]<\displaystyle\frac{2\pi}{3}\right\},
\end{equation*}
with $\arg[k]$ denoting the argument function taking values in the interval $(-2\pi/3,4\pi/3).$ We use $\overline{\Omega_1},$ $\overline{\Omega_2},$
$\overline{\Omega_3},$ $\overline{\Omega_4}$ to denote the closures of the open sectors $\Omega_1,$ $\Omega_2,$ $\Omega_3,$ $\Omega_4,$
respectively, where we recall that the closures are obtained by adding the boundaries to the corresponding open sectors.

There are two relevant particular solutions to \eqref{2.4}, i.e. the left Jost solution $f(k,x)$ and the right Jost solution $g(k,x).$
The left Jost solution $f(k,x)$ is the solution to \eqref{2.4} satisfying the spacial asymptotics as $x\to +\infty$ given by
\begin{equation}
\label{3.10}
\begin{cases}
f(k,x)=e^{kx}\left[1+o(1)\right],\\
\noalign{\medskip}
f'(k,x)=k\,e^{kx}\left[1+o(1)\right],\\
\noalign{\medskip}
f''(k,x)=k^2\,e^{kx}\left[1+o(1)\right],
\end{cases}
\end{equation}
and its $k$-domain is given by $\overline{\Omega_1}.$ The left transmission coefficient $T_{\text{\rm{l}}}(k)$ 
appears in the spacial
asymptotics of $f(k,x)$ when $x\to -\infty,$ where we have
\begin{equation}
\label{3.11}
f(k,x)=e^{kx}\,T_{\text{\rm{l}}}(k)^{-1}[1+o(1)], \qquad k\in \overline{\Omega_1}.
\end{equation}
The right Jost solution $g(k,x)$ is the solution to \eqref{2.4} satisfying the spacial asymptotics as $x\to -\infty$ given by
\begin{equation}
\label{3.12}
\begin{cases}
g(k,x)=e^{kx}\left[1+o(1)\right],\\
\noalign{\medskip}
g'(k,x)=k\,e^{kx}\left[1+o(1)\right],\\
\noalign{\medskip}
g''(k,x)=k^2\,e^{kx}\left[1+o(1)\right],
\end{cases}
\end{equation}
and its $k$-domain is given by $\overline{\Omega_3}.$ The right scattering coefficient $T_{\text{\rm{r}}}(k)$
appears in the spacial
asymptotics of $g(k,x)$ when $x\to +\infty$ as
\begin{equation}
\label{3.13}
g(k,x)=e^{kx}\,T_{\text{\rm{r}}}(k)^{-1}[1+o(1)], \qquad k\in \overline{\Omega_3}.
\end{equation}
In the reflectionless case, the transmission coefficients $T_{\text{\rm{l}}}(k)$ and $T_{\text{\rm{r}}}(k)$
have meromorphic extensions from their respective $k$-domains $\overline{\Omega_1}$ and $\overline{\Omega_3}$ to the entire complex $k$-plane,
and those extensions satisfy \cite{ACTU2025,ACTU2026}
\begin{equation*}
T_{\text{\rm{r}}}(k)= \displaystyle\frac{1}{T_{\text{\rm{l}}}(k)},\qquad k\in\mathbb C.
\end{equation*}

It is known \cite{ACTU2025} that \eqref{2.4} has the particular solutions $m(k,x)$ and $n(k,x)$ with the respective $k$-domains $\overline{\Omega_2}$ and
$\overline{\Omega_4},$ and in the reflectionless case they satisfy the spacial asymptotics given by 
\begin{equation}
\label{3.15}
m(k,x)=e^{kx}\left[1+o(1)\right], \qquad  x\to-\infty, \quad k\in\overline{\Omega_2}, 
\end{equation}
\begin{equation}
\label{3.16}
m(k,x)= e^{kx}\,T_{\text{\rm{l}}}(z^2k)^{-1}\,T_{\text{\rm{r}}}(zk)[1+o(1)],\qquad  x\to+\infty, \quad k\in\overline{\Omega_2}, 
\end{equation}
\begin{equation}
\label{3.17}
n(k,x)=e^{kx}\left[1+o(1)\right], \qquad  x\to-\infty, \quad k\in\overline{\Omega_4},
\end{equation}
\begin{equation}
\label{3.18}
n(k,x)= e^{kx}\,T_{\text{\rm{l}}}(zk)^{-1}\,T_{\text{\rm{r}}}(z^2k)[1+o(1)],\qquad  x\to+\infty, \quad k\in\overline{\Omega_4}, 
\end{equation}
where we recall that $z$ is the cube root of unity appearing in \eqref{3.5}.

For fixed real values of $x$ and $t,$ the large $k$-asymptotics of the basic solutions $f(k,x),$ $g(k,x),$ $m(k,x),$ and $n(k,x)$ are, respectively, given by \cite{ACTU2025,ACTU2026}
\begin{equation}
\label{3.19}
 f(k,x) = e^{kx}\left[1 +\displaystyle\frac{u_1(x)}{k} + \displaystyle\frac{u_2(x)}{k^2} + O\left(\displaystyle\frac{1}{k^3}\right)\right],\qquad k\to\infty  
 \text{\rm{ in }} \overline{\Omega_1},
 \end{equation}
 \begin{equation}
\label{3.20}
 g(k,x) = e^{kx}\left[1 +\displaystyle\frac{v_1(x)}{k} + \displaystyle\frac{v_2(x)}{k^2} + O\left(\displaystyle\frac{1}{k^3}\right)\right],\qquad k\to\infty 
  \text{\rm{ in }}\overline{\Omega_3}, 
 \end{equation}
 \begin{equation}\label{3.21}
m(k,x)=e^{kx}\left[1+O\left(\displaystyle\frac{1}{k}\right)\right], \qquad  k\to\infty  \text{\rm{ in }} \overline{\Omega_2},
\end{equation}
\begin{equation}\label{3.22}
n(k,x)=e^{kx}\left[1+O\left(\displaystyle\frac{1}{k}\right)\right], \qquad  k\to\infty  \text{\rm{ in }} \overline{\Omega_4},
\end{equation}
where we have defined
\begin{equation}\label{3.23}
 u_1(x):= \displaystyle\frac{1}{3}\int_x^{\infty} dy\, Q(y),\qquad x\in\mathbb R,
 \end{equation}
 \begin{equation}\label{3.24}
 u_2(x):= -\displaystyle\frac{1}{3}\int_x^{\infty} dy \left[Q'(y)-P(y)\right] + \displaystyle\frac{1}{18}\left[\int_x^{\infty} dy\, Q(y)\right]^2,
 \qquad x\in\mathbb R,
  \end{equation}
\begin{equation}\label{3.25}
 v_1(x):= -\displaystyle\frac{1}{3}\int_{-\infty}^x dy\, Q(y),\qquad x\in\mathbb R,
 \end{equation}
 \begin{equation}\label{3.26}
 v_2(x):= \displaystyle\frac{1}{3}\int_{-\infty}^x dy \left[Q'(y)-P(y)\right] + \displaystyle\frac{1}{18}\left[\int_{-\infty}^x dy\, Q(y)\right]^2, \qquad x\in\mathbb R.
  \end{equation}
Using \eqref{3.23}--\eqref{3.26}, we express the potentials $Q$ and $P$ in terms of $u_1(x)$ and $u_2(x)$ as
\begin{equation}\label{3.27}
 Q(x)= -3\,\displaystyle\frac{du_1(x)}{dx}, \qquad x\in\mathbb R,
 \end{equation}
 \begin{equation}\label{3.28}
 P(x)= 3\left[u_1(x)\,\displaystyle\frac{du_1(x)}{dx}- \displaystyle\frac{d^2u_1(x)}{dx^2}-\displaystyle\frac{du_2(x)}{dx}\right], \qquad x\in\mathbb R,
 \end{equation}
 or in terms of $v_1(x)$ and $v_2(x)$ as
 \begin{equation}\label{3.29}
 Q(x)= -3\,\displaystyle\frac{dv_1(x)}{dx}, \qquad x\in\mathbb R,
 \end{equation}
 \begin{equation*}
 P(x)= 3\left[v_1(x)\,\displaystyle\frac{dv_1(x)}{dx}- \displaystyle\frac{d^2v_1(x)}{dx^2}-\displaystyle\frac{dv_2(x)}{dx}\right], \qquad x\in\mathbb R.
 \end{equation*}

Since \eqref{2.4} is a linear homogeneous equation, any constant multiple of a solution is also a solution. 
In terms of the solutions $f(k,x)$ in \eqref{3.10} and \eqref{3.11}, the solution $g(k,x)$ in \eqref{3.12} and \eqref{3.13}, the solution $m(k,x)$ in \eqref{3.15} and \eqref{3.16}, and the solution $n(k,x)$ in \eqref{3.17} and \eqref{3.18}, 
we introduce the solution $\Phi_+(k,x)$ to \eqref{2.4}
with the $k$-domain $\overline{\Omega_1}\cup\overline{\Omega_2}$ and the solution $\Phi_-(k,x)$ to \eqref{2.4} with the $k$-domain
$\overline{\Omega_3}\cup\overline{\Omega_4},$ where we have defined
\begin{equation}\label{3.31}
    \Phi_+(k,x) := 
    \begin{cases}
        T_\text{\rm{l}}(k)\, f(k,x), \qquad k\in \overline{\Omega_1},\\
                \noalign{\medskip}
        m(k,x), \qquad k\in \overline{\Omega_2},
    \end{cases}
\end{equation}
\begin{equation}\label{3.32}
    \Phi_-(k,x) := 
    \begin{cases}
        g(k,x), \qquad k\in \overline{\Omega_3},\\
        \noalign{\medskip}
        n(k,x), \qquad k\in \overline{\Omega_4}.
    \end{cases}
\end{equation}

Using the directed half lines $\mathcal L_1$ and $\mathcal L_3$ defined in \eqref{3.1} and \eqref{3.3}, respectively, we obtain the directed full line
$\mathcal L$ via $\mathcal L:=\mathcal L_1\cup (-\mathcal L_3),$ where we recall that
$-\mathcal L_3$ is obtained from $\mathcal L_3$ by changing its direction.
The parametrization of $\mathcal L$ is given by
\begin{equation}
\label{3.33}
\mathcal L:=\{k\in\mathbb C: k=zs \text{\rm{ for }} s\in(-\infty,+\infty)\}.
\end{equation}
The directed line $\mathcal L$ divides the complex $k$-plane into two half planes $\mathcal P^+$ and $\mathcal P^-$ as shown on the right plot of Figure~\ref{Figure1}.
The open left-half complex plane $\mathcal P^+$ and the open right-half complex plane $\mathcal P^-$ can be parametrized as
\begin{equation*}
\mathcal P^+:=\{k\in\mathbb C: k=zs \text{\rm{ for }} s\in\mathbb C^+\},
\end{equation*}
\begin{equation*}
\mathcal P^-:=\{k\in\mathbb C: k=zs \text{\rm{ for }} s\in\mathbb C^-\},
\end{equation*}
where we use $\mathbb C^+$ and $\mathbb C^-$ to denote the upper-half and lower-half complex planes.
We refer to $\mathcal P^+$ as the plus region and refer to $\mathcal P^-$ as the minus region, as indicated on the right plot of Figure~\ref{Figure1}. We use
$\overline{\mathcal P^+}$ and $\overline{\mathcal P^-}$ to denote their closures. Hence, we have $\overline{\mathcal P^+}:=\mathcal P^+\cup\mathcal L$
and $\overline{\mathcal P^-}:=\mathcal P^-\cup\mathcal L.$

In the reflectionless case, it is known \cite{ACTU2025,ACTU2026} that the two 
quantities $T_\text{\rm{l}}(k) f(k,x)$ and $m(k,x)$ become equal to each other when 
$k\in\mathcal L_2$ and that the two quantities $g(k,x)$ and $n(k,x)$ become equal to each other when $k\in\mathcal L_4.$
Furthermore, in the reflectionless case,
$f(k,x),$ $m(k,x),$ $g(k,x),$ and $n(k,x)$ are each analytic in
$k$ in their respective domains
$\Omega_1,$ $\Omega_2,$ $\Omega_3,$ and $\Omega_4,$ respectively, and
they each are continuous in the closures of those respective sectors.
Consequently, in the reflectionless case, we have the following
properties for $\Phi_+(k,x)$ defined in \eqref{3.31} and $\Phi_-(k,x)$ 
defined in \eqref{3.32} for each fixed $x\in\mathbb R.$
The solution $\Phi_-(k,x)$ is
analytic in $k\in\mathcal P^-$ and it is continuous in $k\in\overline{\mathcal P^-}.$
In the reflectionless case, when the left transmission coefficient
$T_\text{\rm{l}}(k)$ has its poles confined to $\Omega_1,$
the solution $\Phi_+(k,x)$ is
meromorphic in $k\in\mathcal P^+$ with
the poles occurring at the poles of $T_\text{\rm{l}}(k)$ in $\Omega_1,$
and it is continuous in $k\in\overline{\mathcal P^+}$
except at those poles.

\section{The Riemann--Hilbert problem related to the $\mathbf N$-soliton solution}
\label{section4} 

Since our main goal in this paper is to obtain the $\mathbf N$-soliton solution to the Sawada--Kotera equation \eqref{1.1} with real-valued $Q$ and $P,$
the corresponding transmission coefficients for \eqref{2.4} must be chosen
in a particular way. We choose the left transmission coefficient $T_{\text{\rm{l}}}(k)$ as
\begin{equation}
\label{4.1}
T_{\text{\rm{l}}}(k)=\displaystyle\frac{\Gamma(k)}{\Gamma(-k)},\qquad k\in\mathbb C,
\end{equation}
where $\Gamma(k)$ is given by
\begin{equation}
\label{4.2}
\Gamma(k):=\displaystyle\prod_{j=1}^{\mathbf N} (k+k_j)(k+k_j^\ast),\qquad k\in\mathbb C,
\end{equation}
by recalling that the asterisk denotes complex conjugation. From \eqref{4.1} and \eqref{4.2} we see that
the poles of 
$T_{\text{\rm{l}}}(k)$ 
are simple and they occur at $k=k_j$ and $k=k_j^\ast$ for
$1\le j\le \mathbf N.$ As indicated in Section~\ref{section3}, we would like to choose
the poles of $T_{\text{\rm{l}}}(k)$ in the sector $\Omega_1$ and we would like
the corresponding potentials $Q$ and $P$ to be real valued.
These requirements impose the necessary but not sufficient restrictions on the location of each $k_j$ so that we must have 
$\arg[k_j]=5\pi/6$ or 
$\arg[k_j]=7\pi/6.$ Since the poles 
occur at $k=k_j$ and $k=k_j^\ast$ for
$1\le j\le \mathbf N,$ there is no loss of generality in using $\arg[k_j]=7\pi/6.$ 
Then, we automatically have
$\arg[k_j^\ast]=5\pi/6.$
Because we only consider simple bound states, the restriction can be presented as
\begin{equation}
\label{4.3}
k_j=iz \eta_j, \qquad 1\le j\le {\mathbf N},
\end{equation}
where each $\eta_j$ is positive and we recall that $z$ is the special complex constant in \eqref{3.5}.
Without loss of generality, we can assume that $0<\eta_1<\dots<\eta_{\mathbf N}.$

From \eqref{4.1} and \eqref{4.2} we see that the left
transmission coefficient $T_{\text{\rm{l}}}(k)$ is defined on the whole complex $k$-plane and that it has $2\mathbf N$ poles located at $k=k_j$ and $k=k_j^\ast,$ respectively, for
$1\le j\le{\mathbf N}.$ 
The large $k$-asymptotics of $T_{\text{\rm{l}}}(k)$ is given by
\begin{equation*}
T_{\text{\rm{l}}}(k)=1+\displaystyle\frac{2\,\Sigma_{\mathbf N}}{k}+\displaystyle\frac{2\,\Sigma_{\mathbf N}^2}{k^2}+O\left(\displaystyle\frac{1}{k^3}\right),\qquad k\to\infty  \text{\rm{ in }}\mathbb C, 
 \end{equation*}
where we have defined
\begin{equation}\label{4.5}
\Sigma_{\mathbf N}:=\sum_{j=1}^{\mathbf N} \left(k_j+k_j^\ast\right).
 \end{equation}
 From \eqref{4.5} we see that $\Sigma_{\mathbf N}$ is a real-valued constant.

With the left transmission coefficient given in \eqref{4.1}, it is known \cite{ACTU2025,ACTU2026} that \eqref{2.4} has bound states at $k=k_j$ and $k=k_j^\ast$ for
$1\le j\le\mathbf N.$ In other words, \eqref{2.4} has a square integrable solution 
at each of those $2\mathbf N$ $k$-values. In fact,
$f(k_j,x)$ and $f(k_j^\ast,x)$ each correspond to a bound-state solution to \eqref{2.4}.
Since \eqref{2.4} is linear and homogeneous, any constant
multiple of $f(k_j,x)$ is also a bound-state solution at $k=k_j$ 
and any constant multiple of $f(k_j^\ast,x)$ is also
a bound-state solution at $k=k_j^\ast.$ It is known \cite{ACTU2025,ACTU2026} that $g(zk_j,x)$ is a bound-state
solution to \eqref{2.4} at $k=k_j$ and that
$g(z^2k_j^\ast,x)$ is a bound-state
solution to \eqref{2.4} at $k=k_j^\ast.$ Thus, there exists a constant $D_j,$ which we call
the bound-state dependency constant at $k=k_j,$ such that 
\begin{equation}\label{4.6}
f(k_j,x)=D_j \, g(zk_j,x),\qquad x\in\mathbb R.
\end{equation}
We recall that we suppress the $t$-dependency in our notation for the potentials $Q$ and $P$ and the 
Jost solutions $f(k,x)$ and $g(k,x).$
It is known \cite{ACTU2025,ACTU2026} that the bound-state dependency constant at $k=k_j^\ast$
in this case is given by $D_j^\ast,$ i.e. we have
\begin{equation}
\label{4.7}
f(k_j^\ast,x)=
D_j^\ast\, g(z^2k_j^\ast,x), \qquad x\in\mathbb R.
\end{equation}

When the potentials $Q$ and $P$ are evolved in time in a way compatible with
the linear operator $A_1$ or $A_2$ defined
in the second lines
of \eqref{2.1} and \eqref{2.2}, respectively,
the time evolution of the dependency constants $D_j$ and $D_j^\ast$ are given by \cite{ACTU2025,ACTU2026}
\begin{equation}\label{4.8}
D_j=E_j\,e^{-9[k_j^5+(k_j^\ast)^5] t},\quad
D_j^\ast=E_j^\ast\,e^{-9[k_j^5+(k_j^\ast)^5] t},
\end{equation}
where $E_j$ and $E_j^\ast$ denote the
values of the dependency constants
$D_j$ and $D_j^\ast,$ respectively, at $t=0.$

In the reflectionless case, it is known \cite{ACTU2025,ACTU2026} that the solution
 $\Phi_+(k,x)$ and the solution  $\Phi_-(k,x)$ coincide on the intersection
 of their respective domains
 $\overline{\mathcal P^+}$ and
  $\overline{\mathcal P^-}.$
  We recall from the right plot in Figure~\ref{Figure1} that the intersection
  of those domains is given by the directed full line $\mathcal L$ defined in \eqref{3.33}.
  Thus, we have
  \begin{equation}\label{4.9}
\Phi_+(k,x)=\Phi_-(k,x), \qquad k\in\mathcal L.
\end{equation}
We remark that \eqref{4.9} constitutes a Riemann--Hilbert problem as follows.
The left transmission coefficient $T_\text{\rm{l}}(k)$ described in \eqref{4.1} and \eqref{4.2} is used
as input to our Riemann--Hilbert problem. For each fixed $x\in\mathbb R,$ we are interested in finding a sectionally
meromorphic function $\Phi(k,x)$ in such a way that $\Phi_+(k,x)$ is the meromorphic section in 
the region $\mathcal P^+$ with the poles coinciding with the poles of
$T_\text{\rm{l}}(k)$ there and that
$\Phi_-(k,x)$ is the analytic section in 
the region $\mathcal P^-.$ 
As seen from \eqref{3.19}--\eqref{3.22}, \eqref{3.31}, and \eqref{3.32},
the sectionally meromorphic function $\Phi(k,x)$ has the behavior
$1+O(1/k)$ as $k\to\infty$ in $\mathbb C.$
As indicated on the right plot of Figure~\ref{Figure1}, it is
appropriate to refer to $\mathcal P^+$ and $\mathcal P^-$ as
the plus and minus regions, respectively, and
refer to $\Phi_+(k,x)$ and $\Phi_-(k,x)$
as the plus and minus functions, respectively.
We refer the reader to \cite{ACTU2025,ATU2025} for the generalization of that Riemann--Hilbert problem
when the reflection coefficients are nonzero.

For the solution of the Riemann--Hilbert problem given in \eqref{4.9}, we refer the reader to
\cite{ACTU2025,ACTU2026}. In order to have a unique solution, it is appropriate to use
the input $\{k_j,E_j\}_{j=1}^{\mathbf N}.$ The solution steps are summarized as follows:

\begin{enumerate}

\item[\text{\rm(a)}] 
We multiplying both sides
of \eqref{4.9} by $e^{-kx}\,\Gamma(-k),$ and we get
\begin{equation}\label{4.10}
e^{-kx}\,\Gamma(-k)\,\Phi_+(k,x)=e^{-kx}\,\Gamma(-k)\,\Phi_-(k,x), \qquad k\in \mathcal L,
\end{equation}
where we recall that $\Gamma(k)$ is the quantity defined in \eqref{4.2}.
For each fixed $x\in\mathbb R,$ from the properties of
$\Phi_+(k,x)$ and $\Phi_-(k,x),$ it follows that
the left-hand side of \eqref{4.10} is analytic in $k\in\mathcal P^+$ and the right-hand side is analytic in
$k\in\mathcal P^-.$ 
As a consequence of the equality in \eqref{4.10},
those two sides are analytic continuations of each other
and that each side is entire
with their respective analytic continuations in $k\in\mathbb C.$ 
Furthermore, using the generalized Liouville theorem \cite{R1987}, we conclude that each side of \eqref{4.10} is equal to a monic
polynomial in $k$ of degree $2\mathbf N,$ where the coefficients may depend on $x$ and $t.$ 
We suppress the $t$-dependence of those coefficients
and write the solution to our Riemann--Hilbert problem as
\begin{equation}\label{4.11}
\Phi_+(k,x)=e^{kx}\,\displaystyle\frac{k^{2\mathbf N}+V(k)\,\mathbf A(x)}{\Gamma(-k)}, \qquad k\in \overline{\mathcal P^+}, \quad x\in\mathbb R,
\end{equation}
\begin{equation}\label{4.12}
\Phi_-(k,x)=e^{kx}\,\displaystyle\frac{k^{2\mathbf N}+V(k)\,\mathbf A(x)}{\Gamma(-k)}, \qquad k\in \overline{\mathcal P^-}, \quad x\in\mathbb R,
\end{equation}
where $V(k)$ is the row vector with the $2\mathbf N$ components defined as
\begin{equation}
\label{4.13}
V(k):=\begin{bmatrix}
k^{2\mathbf N-1} &k^{2\mathbf N-2} & \cdots & k &1
\end{bmatrix},\qquad k\in\mathbb C,
\end{equation}
and $\mathbf A(x)$ is a column vector with the $2\mathbf N$ entries that are functions of $x$ and $t.$
We write $\mathbf A(x)$ in terms of its components as
\begin{equation}
\label{4.14}
\mathbf A(x)=\begin{bmatrix}
A_{2\mathbf N-1}(x)\\
A_{2\mathbf N-2}(x)\\
\vdots\\
A_1(x)\\
A_0(x)
\end{bmatrix},\qquad x\in\mathbb R.
\end{equation}

\item[\text{\rm(b)}] 
Using \eqref{3.31} and \eqref{4.11}, with the help of \eqref{4.1} we obtain
\begin{equation}\label{4.15}
f(k,x)=e^{kx}\,\displaystyle\frac{k^{2\mathbf N}+V(k)\,\mathbf A(x)}{\Gamma(k)}, \qquad k\in\overline{\Omega_1}, \quad x\in \mathbb R.
\end{equation}
Similarly, from \eqref{3.32} and \eqref{4.12} we get
\begin{equation}\label{4.16}
g(k,x)=e^{kx}\,\displaystyle\frac{k^{2\mathbf N}+V(k)\,\mathbf A(x)}{\Gamma(-k)}, \qquad k\in\overline{\Omega_3}, \quad x\in \mathbb R.
\end{equation}
From \eqref{4.15} and \eqref{4.16} we observe that
$f(k,x)$ and $g(k,x)$ have meromorphic extensions from their respective original $k$-domains of analyticity to
the entire complex $k$-plane.

\item[\text{\rm(c)}] 
Using \eqref{4.15} and \eqref{4.16} in \eqref{4.6} and \eqref{4.7}, we obtain a linear
algebraic system of $2\mathbf N$ equations in the $2\mathbf N$ unknowns $A_l(x)$ for
$0\le l\le 2\mathbf N-1.$
With the help of \eqref{4.8}, we write that linear system as
\begin{equation}\label{4.17}
\begin{cases}e^{k_jx}\, \displaystyle\frac{k_j^{2\mathbf N}+V(k_j)\,\mathbf A(x)}{\Gamma(k_j)}
=E_j \,e^{-9[k_j^5+(k_j^\ast)^5]t}\, e^{zk_jx}\,\displaystyle\frac{(zk_j)^{2\mathbf N}+V(zk_j)\,\mathbf A(x)}{\Gamma(-zk_j)},
\\
\noalign{\medskip}
e^{k_j^\ast x}\, \displaystyle\frac{(k_j^\ast)^{2\mathbf N}+V(k_j^\ast)\,\mathbf A(x)}{\Gamma(k_j^\ast)}
=E_j^\ast \,e^{-9[k_j^5+(k_j^\ast)^5]t}\, e^{z^2k_j^\ast x}\,\displaystyle\frac{(z^2 k_j^\ast )^{2\mathbf N}+V(z^2 k_j^\ast)\,\mathbf A(x)}{\Gamma(-z^2 k_j^\ast)},
\end{cases}
\end{equation}
where we have $1\le j\le \mathbf N.$
We remark that the second line in \eqref{4.17} is obtained by taking the complex conjugate of the first line
due to the fact that $\mathbf A(x)$ is real valued and we have
\begin{equation}\label{4.18}
\Gamma(k^\ast)=\Gamma(k)^\ast,
\quad
V(k^\ast)=V(k)^\ast,
\qquad k\in\mathbb C,
\end{equation}
which directly follow from \eqref{4.2} and \eqref{4.13}, respectively.

\item[\text{\rm(d)}] 
We introduce the quantities $\chi_j$ and $\gamma_j$ as
\begin{equation}\label{4.19}
\chi_j:=e^{-(k_j +k_j^\ast)x-9[k_j^5+(k_j^\ast)^5]t}, \qquad 1\le j\le \mathbf N,
\end{equation}
\begin{equation}\label{4.20}
\gamma_j:=-E_j\,\displaystyle\frac{\Gamma(k_j)}{\Gamma(-zk_j)}, \qquad 1\le j\le \mathbf N.
\end{equation}
From \eqref{4.19} it follows that $\chi_j$ is a real-valued function of $x$ and $t$
and that we have $\chi_j\to 0$ as $x\to-\infty$ for each fixed $t.$
From \eqref{4.20} we observe that $\gamma_j$ is a complex-valued constant
directly proportional to the dependency constant $E_j$ at $t=0.$
We refer to $\gamma_j$ as the modified dependency constant at $t=0$ for the bound state at $k=k_j.$
From \eqref{4.8} and the first equality of \eqref{4.18} it follows that
$\gamma_j^\ast$ corresponds to the 
modified dependency constant at $t=0$ for the bound state at $k=k_j^\ast.$
Using \eqref{4.19} and \eqref{4.20} in \eqref{4.17}, we write the linear algebraic system \eqref{4.17} as
\begin{equation}\label{4.21}
\begin{cases}
k_j^{2\mathbf N}+V(k_j)\,\mathbf A(x)=-\gamma_j\,\chi_j\left[(zk_j)^{2\mathbf N}+V(zk_j)\,\mathbf A(x)\right], \qquad 1\le j\le \mathbf N,
\\
\noalign{\medskip}
(k_j^\ast)^{2\mathbf N}+V(k_j^\ast)\,\mathbf A(x)=-\gamma_j^\ast\,\chi_j\left[(z^2k_j^\ast)^{2\mathbf N}
+V(z^2k_j^\ast)\,\mathbf A(x)\right], \qquad 1\le j\le \mathbf N.
\end{cases}
\end{equation}

\item[\text{\rm(e)}] 
The linear algebraic system \eqref{4.21} can be written as
\begin{equation}\label{4.22}
\mathbf M(x)\, \mathbf A(x)=-\mathbf B(x),
\end{equation}
where we have defined the $2\mathbf N\times 2\mathbf N$ matrix $\mathbf M(x)$ and the column vector $\mathbf B(x)$ with $2\mathbf N$ entries as
\begin{equation}\label{4.23}
\mathbf M(x):=\begin{bmatrix}
m_{2\mathbf N-1}(k_1) & m_{2\mathbf N-2}(k_1) & \cdots & m_1(k_1) & m_0(k_1)\\
m_{2\mathbf N-1}(k_1^\ast) & m_{2\mathbf N-2}(k_1^\ast) & \cdots & m_1(k_1^\ast) & m_0(k_1^\ast)\\
\vdots & \vdots & \ddots & \vdots & \vdots \\
m_{2\mathbf N-1}(k_{\mathbf N}) & m_{2\mathbf N-2}(k_{\mathbf N}) & \cdots & m_1(k_{\mathbf N}) & m_0(k_{\mathbf N})\\
m_{2\mathbf N-1}(k_{\mathbf N}^\ast) & m_{2\mathbf N-2}(k_{\mathbf N}^\ast) & \cdots & m_1(k_{\mathbf N}^\ast) & m_0(k_{\mathbf N}^\ast)
\end{bmatrix},
\end{equation}
\begin{equation}\label{4.24}
\mathbf B(x):=\begin{bmatrix}
m_{2\mathbf N}(k_1)\\
m_{2\mathbf N}(k_1^\ast)\\
\vdots\\
m_{2\mathbf N}(k_{\mathbf N})\\
m_{2\mathbf N}(k_{\mathbf N}^\ast)
\end{bmatrix},
\end{equation}
with the entries in \eqref{4.23} and \eqref{4.24} defined as
\begin{equation}\label{4.25}
\begin{cases}
m_l(k_j):=k_j^l+(zk_j)^l\,\gamma_j\,\chi_j,
\qquad 1\le j\le \mathbf N, \quad 0\le l\le 2\mathbf N,
\\
\noalign{\medskip}
m_l(k_j^\ast):=(k_j^\ast)^l+(z^2k_j^\ast)^l\,\gamma_j^\ast\,\chi_j,\qquad 1\le j\le \mathbf N, \quad 0\le l\le 2\mathbf N.
\end{cases}
\end{equation}
From \eqref{4.25} we observe that
\begin{equation}
\label{4.26}
m_l(k_j^\ast)=m_l(k_j)^\ast,
\qquad 1\le j\le \mathbf N, \quad 0\le l\le 2\mathbf N.
\end{equation}
We remark that we have suppressed the $t$-dependence in our notation for $\mathbf M(x),$ $\mathbf B(x),$ and $m_l(k_j).$ In fact, the only dependence on $x$ and $t$ in those three quantities is through the
scalar function $\chi_j$ for $1\le j\le \mathbf N.$

\item[\text{\rm(f)}] 
We recall that the input to our Riemann--Hilbert problem \eqref{4.9} is $\{k_j,E_j\}_{j=1}^{\mathbf N}.$
From \eqref{4.2} and \eqref{4.20} it follows that our input data set 
is equivalent to 
$\{k_j,\gamma_j\}_{j=1}^{\mathbf N}.$
From \eqref{4.23}--\eqref{4.26} we see that $\mathbf M(x)$ and $\mathbf B(x)$ are each uniquely
determined by our input data set.
We solve \eqref{4.22} and obtain the column vector $\mathbf A(x)$ in \eqref{4.14} as
\begin{equation}\label{4.27}
\mathbf A(x)=-\mathbf M(x)^{-1}\mathbf B(x).
\end{equation}
We refer the reader to \cite{ACTU2025,ACTU2026} for further details and other equivalent expressions for
$\mathbf A(x)$ given in terms of the input data set $\{k_j,\gamma_j\}_{j=1}^{\mathbf N}.$
From \eqref{3.19}, \eqref{3.27}, \eqref{3.28}, and \eqref{4.15}
it follows that the potential $Q$ is determined
by $A_{2\mathbf N-1}(x)$ alone and the potential $P$ is determined by $A_{2\mathbf N-1}(x)$ and $A_{2\mathbf N-2}(x)$ only. The remaining entries 
$A_j(x)$ for $0\le j \le 2\mathbf N-3$
are used only to determine the solutions to the third-order equation \eqref{2.4}.
In the special case when $\mathbf N=1,$ we only have $A_1(x)$ and $A_0(x).$

\item[\text{\rm(g)}] 
Having determined $A_{2\mathbf N-1}(x)$ and $A_{2\mathbf N-2}(x)$ via \eqref{4.27} by using the input data set 
$\{k_j,\gamma_j\}_{j=1}^{\mathbf N},$ we next 
determine the potentials $Q$ and $P$ in \eqref{2.4}. Using \eqref{4.2}, \eqref{4.13}, \eqref{4.15}, we obtain 
the asymptotics of $f(k,x)$ when $k\to \infty$ in $\mathbb C$ 
as 
\begin{equation}\label{4.28}
f(k,x)
=e^{kx}\left[1+\displaystyle\frac{A_{2\mathbf N-1}(x)-\Sigma_{\mathbf N}}{k}
   +\displaystyle\frac{A_{2\mathbf N-2}(x)-\Sigma_{\mathbf N} A_{2\mathbf N-1}(x)+\Pi_{\mathbf N}}{k^2}+O\left(\displaystyle\frac{1}{k^3}\right)\right],
\end{equation}
where $\Sigma_{\mathbf N}$ is the constant in \eqref{4.5} and
$\Pi_{\mathbf N}$ is the constant defined
as
\begin{equation}\label{4.29}
\begin{split}
\Pi_{\mathbf N}:=&k_1\left(k_1+k_1^\ast+\cdots+k_{\mathbf N}+k_{\mathbf N}^\ast \right)+k_1^\ast\left(k_1^\ast+k_2+\cdots+k_{\mathbf N}+k_{\mathbf N}^\ast \right)\\
       &+k_2\left(k_2+k_2^\ast+\cdots+k_{\mathbf N}+k_{\mathbf N}^\ast \right)+\cdots
       +k_{\mathbf N}\left(k_{\mathbf N}+k_{\mathbf N}^\ast\right)+ k_{\mathbf N}^\ast\left(k_{\mathbf N}^\ast\right).
\end{split}
\end{equation}
Comparing \eqref{3.19} with \eqref{4.28} we get
\begin{equation}\label{4.30}
u_1(x)=A_{2\mathbf N-1}(x)-\Sigma_{\mathbf N},
\end{equation}
\begin{equation}\label{4.31}
u_2(x)=A_{2\mathbf N-2}(x)-\Sigma_{\mathbf N} A_{2\mathbf N-1}(x)+\Pi_{\mathbf N}.
\end{equation}
With the help of \eqref{3.27}, \eqref{3.28},  \eqref{4.30} and \eqref{4.31}, we recover
$Q$ and $P$ appearing in \eqref{2.4} as
\begin{equation}\label{4.32}
Q(x)=-3\,\displaystyle\frac{dA_{2\mathbf N-1}(x)}{dx},
\end{equation}
\begin{equation}\label{4.33}
P(x)=3\left(A_{2\mathbf N-1}(x)\,\displaystyle\frac{dA_{2\mathbf N-1}(x)}{dx}-\displaystyle\frac{d^2A_{2\mathbf N-1}(x)}{dx^2}-\displaystyle\frac{dA_{2\mathbf N-2}(x)}{dx}\right).
\end{equation}
Since the potentials $Q$ and $P$ vanish as $x\to+\infty,$ it follows from \eqref{3.23} and \eqref{3.24} that
$u_1(x)$ and $u_2(x)$ each vanish as $x\to+\infty.$
Consequently, from \eqref{4.30} and \eqref{4.31} we get
\begin{equation}\label{4.34}
\displaystyle\lim_{x\to+\infty} A_{2\mathbf N-1}(x)=\Sigma_{\mathbf N},\quad
\displaystyle\lim_{x\to+\infty} A'_{2\mathbf N-1}(x)=0,\quad
\displaystyle\lim_{x\to+\infty} A_{2\mathbf N-2}(x)=\Sigma_{\mathbf N}^2-\Pi_{\mathbf N}.
\end{equation}

\end{enumerate}

\section{The $\mathbf N$-soliton solution to the Sawada--Kotera equation}
\label{section5} 

Our goal in this section is to provide an explicit expression for the $\mathbf N$-soliton solution to the Sawada--Kotera equation \eqref{1.1}. We achieve
our goal by solving the inverse scattering problem for the linear equation \eqref{2.4} in the reflectionless case with the input data set consisting of the
bound-state poles of the transmission coefficient $T_{\text{\rm{l}}}(k)$ in \eqref{4.1} and the corresponding bound-state dependency constants. The reason for the choice
of the $k_j$-values given in \eqref{4.3} is due to the fact that the soliton solution $Q$ must be real valued. There is a further restriction that we must impose on our solution to the inverse scattering problem, namely we must have $P\equiv 0$ in the SK1 case and have $P=Q_x$ in the SK2 case.

In this section we show that the appropriate restrictions in the SK1 and SK2 cases can be imposed by using 
certain constraints on the complex-valued
modified bound-state dependency constants $\gamma_j$ appearing in \eqref{4.20}. A lengthy
initial part of our section is devoted to derive the appropriate restrictions on
the modified dependency constants $\gamma_j.$ Those restrictions amount
to expressing the ratio of the imaginary part to the real part of each $\gamma_j$ in terms of
the bound-state parameters $k_1,$
$k_2,$
$\dots,$
$k_{\mathbf N}$ appearing in \eqref{4.3}. Consequently, we express the $\mathbf N$-soliton solution $Q$ to \eqref{1.1}
by using the 
parameters $k_j$ and the real parts of $\gamma_j$
for $1\le j \le \mathbf N.$ 
Later, we replace each real part of $\gamma_j$ by the parameter $c_j$
by using \eqref{5.42} and \eqref{5.43} in the 
SK1 and SK2 cases, respectively, for $1\le j \le \mathbf N.$ We then express the $\mathbf N$-soliton solution to the
SK equation in terms of the $2\mathbf N$ real parameters
in the set $\{\eta_j,c_j\}_{j=1}^\mathbf N,$ where the parameters $\eta_j$ are distinct and positive. The advantage of using $c_j$ instead of the real part of $\gamma_j$ in our input data set is the following.
The function $Q$ does not contain any singularity
when $x$ and $t$ take all values in $\mathbb R$ if each parameter $c_j$ takes any positive value,
whereas to avoid any singularity in
$Q,$ the real part of $\gamma_j$ may need to be restricted to positive or negative values or to a constrained range.

In Section~\ref{section4}, we have solved the inverse scattering problem for \eqref{2.4} in the reflectionless case
by using the input data set $\{k_j,E_j\}_{j=1}^{\mathbf N}$ or the equivalent data set
$\{k_j,\gamma_j\}_{j=1}^{\mathbf N}.$
We have recovered the Jost solutions $f(k,x)$ and $g(k,x)$ and the 
potentials $Q$ and $P$ as in \eqref{4.15}, \eqref{4.16},
\eqref{4.32}, \eqref{4.33}, respectively,
where $\mathbf A(x)$ is expressed in terms of our input data set via \eqref{4.27}.
In the determination of each of $A_{2\mathbf N-1}(x)$ and
$A_{2\mathbf N-2}(x),$ we have used the modified dependency constants $\gamma_j$ and
$\gamma_j^\ast$ for $1\le j\le \mathbf N.$
In order for $Q$ to satisfy the Sawada--Kotera equation \eqref{1.1},
as indicated in Section~\ref{section2},
the potential $P$ must satisfy
the restriction $P\equiv 0$ in the SK1 case and the restriction
$P=Q_x$ in the SK2 case. 

Let us use $r_j$ and $s_j$ to denote the real and imaginary parts of the
complex-valued $\gamma_j$ so that we have
\begin{equation}\label{5.1}
\gamma_j=r_j\left(1+i\,\displaystyle\frac{s_j}{r_j}\right),
\quad
\gamma_j^\ast=r_j\left(1-i\,\displaystyle\frac{s_j}{r_j}\right),
\qquad 1\le j\le \mathbf N.
\end{equation}
Our motivation behind separating the complex-valued dependency constant
$\gamma_j$ into its real and imaginary parts $r_j$ and $s_j,$ respectively, can be
explained as follows. The $\mathbf N$-soliton solution
$Q$ to the Sawada--Kotera equation \eqref{1.1} is a real-valued function of $x$ and $t$ containing the parameters
related to the bound-state poles and the bound-state dependency constants. Hence, it is natural that we use real-valued parameters when we express the
real-valued solution $Q$ to \eqref{1.1}. From \eqref{4.3} we see that we can use the positive parameter $\eta_j$ instead of the 
complex-valued parameter $k_j$ for
$1\le j \le \mathbf N.$ Similarly, instead of using the complex-valued parameter $\gamma_j,$ we can use the real parameters $r_j$ and $s_j$
for $1\le j\le\mathbf N.$ 
 It turns out that the two aforementioned restrictions in the SK1 and SK2 cases comprise the
 specification of the ratios $s_j/r_j$ for $1\le j\le \mathbf N$ in each case.
 Consequently, the $\mathbf N$-soliton solution to the Sawada--Kotera equation
 \eqref{1.1} is determined by the input data set $\{k_j,r_j\}_{j=1}^{\mathbf N}$
 when the ratios $s_j/r_j$ for $1\le j\le \mathbf N$ are specified.
 Since the $\mathbf N$ constants $k_j$ appear symmetrically in the input data set,
 without loss of generality it is sufficient to determine the ratio
 $s_1/r_1.$ Then, using the symmetry we can get the remaining ratios
 $s_j/r_j$ for $2\le j\le \mathbf N.$ We have the explicit formula for the ratio
 $s_1/r_1$ presented in (6.14) of \cite{ACTU2026} covering both the SK1 and SK2 cases.
 In this section, we use a slightly different method, but still a general method, to determine
 the ratios $s_1/r_1$ in each of the SK1 and SK2 cases.

We know from \eqref{4.32} and \eqref{4.33} that either of the two aforementioned 
restrictions in the cases of SK1 and SK2, respectively, involves the corresponding restrictions
on $A_{2\mathbf N-1}(x)$ and
$A_{2\mathbf N-2}(x)$ only.
By using Cramer's rule, from \eqref{4.22} we get
\begin{equation}\label{5.2}
A_{2\mathbf N-1}(x)=-\displaystyle\frac{\det[\mathbf M_1(x)]}{\det[\mathbf M(x)]},
\end{equation}
\begin{equation}\label{5.3}
A_{2\mathbf N-2}(x)=-\displaystyle\frac{\det[\mathbf M_2(x)]}{\det[\mathbf M(x)]},
\end{equation}
where $\mathbf M_1(x)$ is the $2\mathbf N\times 2\mathbf N$ matrix obtained by replacing the first column of the matrix $\mathbf M(x)$ with the
column vector $\mathbf B(x)$ and that $\mathbf M_2(x)$ denotes the $2\mathbf N\times 2\mathbf N$ matrix obtained by replacing the second column of
$\mathbf M(x)$ with $\mathbf B(x).$
It is known \cite{ACTU2025,ACTU2026} that the determinant of $\mathbf M_1(x)$ can be expressed in terms of the determinant of
$\mathbf M(x)$ and the $x$-derivative of $\det[\mathbf M(x)]$ as
\begin{equation}\label{5.4}
\det[\mathbf M_1(x)]=\Sigma_{\mathbf N}\det[\mathbf M(x)]+\displaystyle\frac{d\det[\mathbf M(x)]}{dx},
\end{equation}
where we recall that $\Sigma_{\mathbf N}$ is the constant defined in \eqref{4.5}.
Using \eqref{5.4} in \eqref{5.2}, we get
\begin{equation}\label{5.5}
A_{2\mathbf N-1}(x)=-
\Sigma_{\mathbf N}-\displaystyle\frac{1}{\det[\mathbf M(x)]}\displaystyle\frac{d\det[\mathbf M(x)]}{dx},
\end{equation}
and then, using \eqref{5.5} in \eqref{4.32}, we express
the potential $Q$ in terms of 
 the determinant of
$\mathbf M(x)$ as
\begin{equation}\label{5.6}
Q(x)=3\,\displaystyle\frac{d}{dx}\left(\displaystyle\frac{1}
{\det[\mathbf M(x)]}\,
\displaystyle\frac{d\det[\mathbf M(x)]}{dx}\right).
\end{equation}

From \eqref{4.23} and \eqref{4.25} we see that the quantity
$\det[\mathbf M(x)]$
 is a polynomial of degree $2\mathbf N$ in the $2\mathbf N$ variables
$\gamma_1\chi_1,$ 
$\gamma_1^\ast\chi_1,$ $\dots,$
$\gamma_{\mathbf N}\chi_{\mathbf N},$
$\gamma^\ast_{\mathbf N}\chi_{\mathbf N},$
where each of those $2\mathbf N$ variables appears at most to the first power.
The appearance of each $\chi_j$ in
$\det[\mathbf M(x)]$
is symmetrical. By letting
$\chi_j=0$ for $2\le j\le \mathbf N,$ from \eqref{4.23} we obtain
\begin{equation}\label{5.7}
\det[\mathbf M(x)]=\det\begin{bmatrix}
k_1^{2\mathbf N-1}+(zk_1)^{2\mathbf N-1}\gamma_1\,\chi_1 &  \cdots &k_1+zk_1\gamma_1\,\chi_1 &1+\gamma_1\,\chi_1\\
\noalign{\medskip}
(k_1^\ast)^{2\mathbf N-1}+(z^2k_1^\ast)^{2\mathbf N-1}\gamma_1^\ast\,\chi_1& \cdots & k_1^\ast +z^2 k_1^\ast\gamma_1^\ast\chi_1& 1+\gamma_1^\ast\chi_1\\
\noalign{\medskip}
\vdots & \ddots & \vdots & \vdots \\
\noalign{\medskip}
k_{\mathbf N} ^{2\mathbf N-1}&  \cdots & k_{\mathbf N}& 1\\
\noalign{\medskip}
(k_{\mathbf N}^\ast) ^{2\mathbf N-1}& \cdots & k_{\mathbf N}^\ast & 1
\end{bmatrix}.
\end{equation}
Let us write \eqref{5.7} as
\begin{equation}\label{5.8}
\det[\mathbf M(x)]=\nu+[\nu_{10}\,\gamma_1+\nu_{01}\,\gamma_1^\ast]\,\chi_1+\nu_{11}\,\gamma_1\,\gamma_1^\ast\chi_1^2,
\end{equation}
where the coefficients $\nu,$ $\nu_{10},$ $\nu_{01},$ and $\nu_{11}$ 
are
explicitly expressed in terms of the 
parameters in the set
$\{k_j,k_j^\ast\}_{j=1}^{\mathbf N}.$
The coefficient $\nu$ is obtained by letting $\chi_1=0$ in \eqref{5.7}, from which
we have
\begin{equation}\label{5.9}
\nu:=\det\begin{bmatrix}
k_1^{2\mathbf N-1} & k_1^{2\mathbf N-2}& \cdots &k_1 &1\\
\noalign{\medskip}
(k_1^\ast)^{2\mathbf N-1}&(k_1^\ast)^{2\mathbf N-2} & \cdots & k_1^\ast & 1\\
\noalign{\medskip}
\vdots & \vdots & \ddots & \vdots & \vdots \\
\noalign{\medskip}
k_{\mathbf N} ^{2\mathbf N-1}& k_{\mathbf N} ^{2\mathbf N-2}& \cdots & k_{\mathbf N}& 1\\
\noalign{\medskip}
(k_{\mathbf N}^\ast) ^{2\mathbf N-1}& (k_{\mathbf N}^\ast) ^{2\mathbf N-2}& \cdots & k_{\mathbf N}^\ast & 1
\end{bmatrix}.
\end{equation}
The right-hand side of \eqref{5.9} can be expressed as a Vandermonde coefficient as an ordered product, as
indicated in (5.49) and (5.50) of \cite{ACTU2026}.
We have
\begin{equation}\label{5.10}
\nu=[(k_1-k_1^\ast)(k_1-k_2)\cdots(k_1-k^\ast_{\mathbf N})][(k^\ast_1-k_2)\cdots(k^\ast_1-k^\ast_{\mathbf N})]
\cdots[(k_{\mathbf N}-k^\ast_{\mathbf N})].
\end{equation}
The coefficient $\nu_{10}$ is obtained from $\nu$ by replacing $k_1$ with $zk_1,$ 
the coefficient $\nu_{01}$ is obtained from $\nu$ by replacing $k^\ast_1$ with $z^2k^\ast_1,$ 
and the coefficient $\nu_{11}$ is obtained from $\nu$ by replacing $k_1$ with $zk_1$ and $k^\ast_1$ with $z^2k^\ast_1.$ 
In other words, we have
\begin{equation}\label{5.11}
\nu_{10}:=\nu\bigg|_{k_1\mapsto zk_1},\quad
\nu_{01}:=\nu\bigg|_{k^\ast_1\mapsto z^2k^\ast_1},\quad
\nu_{11}:=\nu\bigg|_{\substack{k_1\mapsto zk_1\\
k^\ast_1\mapsto z^2k^\ast_1}}.
\end{equation}

We remark that the variable $x$ appears on the right-hand side of \eqref{5.8} only through $\chi_1.$
Thus, when we let $\chi_j=0$ for $2\le j\le \mathbf N,$ from \eqref{4.19} and \eqref{5.8} we obtain
\begin{equation}\label{5.12}
\displaystyle\frac{d\det[\mathbf M(x)]}{dx}=
-(k_1+k^\ast_1)\, [\nu_{10}\,\gamma_1+\nu_{01}\,\gamma_1^\ast]\,\chi_1-2(k_1+k^\ast_1)\,\nu_{11}\,\gamma_1\,\gamma_1^\ast\chi_1^2,
\end{equation}
\begin{equation}\label{5.13}
\displaystyle\frac{d^2\det[\mathbf M(x)]}{dx^2}=
(k_1+k^\ast_1)^2\, [\nu_{10}\,\gamma_1+\nu_{01}\,\gamma_1^\ast]\,\chi_1+4(k_1+k^\ast_1)^2\,\nu_{11}\,\gamma_1\,\gamma_1^\ast\chi_1^2.
\end{equation}

We recall that the matrix $\mathbf M_2(x)$ appearing in \eqref{5.3}
is obtained from the matrix
$\mathbf M(x)$ in \eqref{4.23} by replacing the second column by the vector
$\mathbf B(x)$ in \eqref{4.24}. Consequently, 
$\det[\mathbf M_2(x)]$
 is a polynomial of degree $2\mathbf N$ in the $2\mathbf N$ variables
$\gamma_1\chi_1,$ 
$\gamma_1^\ast\chi_1,$ $\dots,$
$\gamma_{\mathbf N}\chi_{\mathbf N},$
$\gamma^\ast_{\mathbf N}\chi_{\mathbf N},$
where each of those $2\mathbf N$ variables appears at most to the first power.
The appearance of each $\chi_j$ in
$\det[\mathbf M_2(x)]$
is symmetrical. By letting
$\chi_j=0$ for $2\le j\le \mathbf N,$ from \eqref{4.23} we obtain
the reduced form of the matrix with the determinant given by
\begin{equation}\label{5.14}
\det[\mathbf M_2(x)]=\det\begin{bmatrix}
k_1^{2\mathbf N-1}+(zk_1)^{2\mathbf N-1}\gamma_1\,\chi_1 
& k_1^{2\mathbf N}+(zk_1)^{2\mathbf N}\gamma_1\,\chi_1& \cdots &1+\gamma_1\,\chi_1\\
\noalign{\medskip}
(k_1^\ast)^{2\mathbf N-1}+(z^2k_1^\ast)^{2\mathbf N-1}\gamma_1^\ast\,\chi_1& (k_1^\ast)^{2\mathbf N}+(z^2k_1^\ast)^{2\mathbf N}\gamma_1^\ast\,\chi_1 & \cdots & 1+\gamma_1^\ast\,\chi_1\\
\noalign{\medskip}
\vdots & \vdots & \ddots & \vdots \\
\noalign{\medskip}
k_{\mathbf N} ^{2\mathbf N-1}&  k_{\mathbf N} ^{2\mathbf N}&\cdots &  1\\
\noalign{\medskip}
(k_{\mathbf N}^\ast) ^{2\mathbf N-1}& (k_{\mathbf N}^\ast) ^{2\mathbf N}&\cdots & 1
\end{bmatrix}.
\end{equation}
In analogy with \eqref{5.8}, we write \eqref{5.14} as
\begin{equation}\label{5.15}
\det[\mathbf M_2(x)]=\mu+[\mu_{10}\,\gamma_1+\mu_{01}\,\gamma_1^\ast]\,\chi_1+\mu_{11}\,\gamma_1\,\gamma_1^\ast\chi_1^2,
\end{equation}
where the coefficients $\mu,$ $\mu_{10},$ $\mu_{01},$ and $\mu_{11}$ 
are
explicitly expressed in terms of the 
parameters in the set
$\{k_j,k_j^\ast\}_{j=1}^{\mathbf N}.$
The coefficient $\mu$ is obtained by letting $\chi_1=0$ in \eqref{5.7}, from which
we have
\begin{equation}\label{5.16}
\mu:=\det\begin{bmatrix}
k_1^{2\mathbf N-1} & k_1^{2\mathbf N}& \cdots &k_1 &1\\
\noalign{\medskip}
(k_1^\ast)^{2\mathbf N-1}&(k_1^\ast)^{2\mathbf N} & \cdots & k_1^\ast & 1\\
\noalign{\medskip}
\vdots & \vdots & \ddots & \vdots & \vdots \\
\noalign{\medskip}
k_{\mathbf N} ^{2\mathbf N-1}& k_{\mathbf N} ^{2\mathbf N}& \cdots & k_{\mathbf N}& 1\\
\noalign{\medskip}
(k_{\mathbf N}^\ast) ^{2\mathbf N-1}& (k_{\mathbf N}^\ast) ^{2\mathbf N}& \cdots & k_{\mathbf N}^\ast & 1
\end{bmatrix}.
\end{equation}
Using \eqref{4.5}, \eqref{4.29}, and \eqref{5.16} we write $\mu$ in terms of the quantity
$\nu$ in \eqref{5.10} as
\begin{equation}
\label{5.17}
\mu=(\Pi_{\mathbf N}-\Sigma^2_{\mathbf N})\,\nu.
\end{equation}
The coefficient $\mu_{10}$ is obtained from $\mu$ in \eqref{5.17} by replacing $k_1$ with $zk_1,$ 
the coefficient $\mu_{01}$ is obtained from $\mu$ by replacing $k^\ast_1$ with $z^2k^\ast_1,$ 
and the coefficient $\mu_{11}$ is obtained from $\mu$ by replacing $k_1$ with $zk_1$ and $k^\ast_1$ with $z^2k^\ast_1.$ 
In other words, we have
\begin{equation}\label{5.18}
\mu_{10}:=\mu\bigg|_{k_1\mapsto zk_1},\quad
\mu_{01}:=\mu\bigg|_{k^\ast_1\mapsto z^2k^\ast_1},\quad
\mu_{11}:=\mu\bigg|_{\substack{k_1\mapsto zk_1\\
k^\ast_1\mapsto z^2k^\ast_1}}.
\end{equation}

In order to determine the ratio $s_1/r_1$ in the case of SK1, we proceed as follows.
In this case, since $P\equiv0,$ any antiderivative of the right-hand side of \eqref{4.33} must be equal
to a constant and that constant must be equal to the value of the chosen antiderivative at $x=+\infty.$
Hence, with the help of \eqref{4.34}, we integrate \eqref{4.33} and obtain
\begin{equation}\label{5.19}
\displaystyle\frac{1}{2}\left[A_{2\mathbf N-1}(x)\right]^2 -A'_{2\mathbf N-1}(x)-
A_{2\mathbf N-2}(x)=\Pi_{\mathbf N}-\displaystyle\frac{1}{2}\,\Sigma_{\mathbf N}^2,
\end{equation}
where we recall that the prime denotes the $x$-derivative.
Using \eqref{5.3} and \eqref{5.5} in \eqref{5.19}, after some simplification, we obtain
the equivalent expression given by
\begin{equation}\label{5.20}
\begin{split}
\left(\Sigma_{\mathbf N}^2-\Pi_{\mathbf N}
\right)\det[\mathbf M(x)]+\Sigma_{\mathbf N}
\,\displaystyle\frac{d \det[\mathbf M(x)]}{dx}+
&\displaystyle\frac{d^2\det[\mathbf M(x)]}{dx^2}+
\det[\mathbf M_2(x)]\\
&=\displaystyle\frac{1}{
2\,\det[\mathbf M(x)]}\left(
\displaystyle\frac{d \det[\mathbf M(x)]}{dx}
\right)^2.
\end{split}
\end{equation}
Multiplying both sides of \eqref{5.20} by $2\det[\mathbf M(x)],$
we obtain the equivalent equality given by
\begin{equation}\label{5.21}
\begin{split}
2 \left(\Sigma_{\mathbf N}^2-\Pi_{\mathbf N}
\right)\left(\det[\mathbf M(x)]\right)^2+2\, \Sigma_{\mathbf N} &\det[\mathbf M(x)]
\,\displaystyle\frac{d \det[\mathbf M(x)]}{dx}+
2\det[\mathbf M(x)]\,\displaystyle\frac{d^2\det[\mathbf M(x)]}{dx^2}\\
&+2\det[\mathbf M(x)]\,
\det[\mathbf M_2(x)]
-\left(
\displaystyle\frac{d \det[\mathbf M(x)]}{dx}
\right)^2=0.
\end{split}
\end{equation}
If we let $\chi_j=0$ for $2\le j\le \mathbf N$ in \eqref{5.21}, we see that 
the left-hand side of \eqref{5.21} becomes a polynomial  of degree four in $\chi_1.$
Using \eqref{5.8}, \eqref{5.12}, \eqref{5.13}, and \eqref{5.15} in \eqref{5.21} we observe that
the equality in \eqref{5.21} is satisfied if and only if the coefficient of $\chi_1$ on the left-hand side is chosen as zero.
Using \eqref{5.1}, we see that the aforementioned coefficient vanishes if and only if we have
\begin{equation}\label{5.22}
\displaystyle\frac{s_1}{r_1}=i\,\displaystyle\frac{\alpha_1\,(\nu_{10}+\nu_{01})+(\mu_{10}+\mu_{01})
}{\alpha_1\,(\nu_{10}-\nu_{01})+(\mu_{10}-\mu_{01})},
\end{equation}
where we have defined
\begin{equation}\label{5.23}
\alpha_1:=(\Sigma^2_{\mathbf N}-\Pi_{\mathbf N})-(k_1+k^\ast_1)\,\Sigma_{\mathbf N}+
(k_1+k^\ast_1)^2.
\end{equation}
Thus, we have shown that the ratio $s_1/r_1$ in the case of SK1 is uniquely determined by the parameters
in the set $\{k_j,k_j^\ast\}_{j=1}^{\mathbf N}.$

To determine the ratio $s_1/r_1$ in terms of the parameters
in the set $\{k_j,k_j^\ast\}_{j=1}^{\mathbf N}$ in the case of SK2, we proceed as follows.
This time, we have $P=Q_x$ or equivalently we have $P-Q_x\equiv 0.$ Using \eqref{4.32} and \eqref{4.33} 
we see that the latter equality is equivalent to
\begin{equation}\label{5.24}
A_{2\mathbf N-1}(x)\,A'_{2\mathbf N-1}(x)-A'_{2\mathbf N-2}(x)=0.
\end{equation}
Integrating \eqref{5.24} and using \eqref{4.34} we get
\begin{equation}\label{5.25}
\displaystyle\frac{1}{2}\left[A_{2\mathbf N-1}(x)\right]^2 -
A_{2\mathbf N-2}(x)=\Pi_{\mathbf N}-\displaystyle\frac{1}{2}\,\Sigma_{\mathbf N}^2.
\end{equation}
Using \eqref{5.3} and \eqref{5.5} in \eqref{5.25}, after some simplification we obtain
the equivalent expression
\begin{equation}\label{5.26}
\left(\Sigma_{\mathbf N}^2-\Pi_{\mathbf N}
\right)\det[\mathbf M(x)]+\Sigma_{\mathbf N}
\,\displaystyle\frac{d \det[\mathbf M(x)]}{dx}+
\det[\mathbf M_2(x)]
=-\displaystyle\frac{1}{
2\,\det[\mathbf M(x)]}\left(
\displaystyle\frac{d \det[\mathbf M(x)]}{dx}
\right)^2.
\end{equation}
We remark that \eqref{5.26} is the counterpart in the SK2 case for \eqref{5.20} given
in the case of SK1.
Multiplying both sides of \eqref{5.26} by $2\det[\mathbf M(x)],$
we obtain the equivalent equality given by
\begin{equation}\label{5.27}
\begin{split}
2 \left(\Sigma_{\mathbf N}^2-\Pi_{\mathbf N}
\right)\left(\det[\mathbf M(x)]\right)^2+&2 \Sigma_{\mathbf N}\,\det[\mathbf M(x)]\,
\displaystyle\frac{d \det[\mathbf M(x)]}{dx}
\\
&+2\det[\mathbf M(x)]\,
\det[\mathbf M_2(x)]
+\left(
\displaystyle\frac{d \det[\mathbf M(x)]}{dx}
\right)^2=0.
\end{split}
\end{equation}
If we let $\chi_j=0$ for $2\le j\le \mathbf N$ in \eqref{5.27}, we observe that 
the left-hand side of \eqref{5.27} becomes a polynomial of degree four in $\chi_1.$ 
Using \eqref{5.8}, \eqref{5.12}, \eqref{5.13}, and \eqref{5.15} in \eqref{5.27}, we see that
the equality in \eqref{5.27} is satisfied if and only if the coefficient of $\chi_1$ on the left-hand side is chosen as zero.
Using \eqref{5.1}, we see that the aforementioned coefficient vanishes if and only if we have
\begin{equation}\label{5.28}
\displaystyle\frac{s_1}{r_1}=i\,\displaystyle\frac{\beta_1\,(\nu_{10}+\nu_{01})+(\mu_{10}+\mu_{01})
}{\beta_1\,(\nu_{10}-\nu_{01})+(\mu_{10}-\mu_{01})},
\end{equation}
where we have defined
\begin{equation}\label{5.29}
\beta_1:=(\Sigma^2_{\mathbf N}-\Pi_{\mathbf N})-(k_1+k^\ast_1)\,\Sigma_{\mathbf N}.
\end{equation}
Comparing \eqref{5.23} and \eqref{5.29}, we see that $\alpha_1$ and $\beta_1$ differ from each other by only one term.
Thus, we have shown that the ratio $s_1/r_1$ in the case of SK2 is uniquely determined by the parameters
in the set $\{k_j,k_j^\ast\}_{j=1}^{\mathbf N}.$

As we have already indicated, because of the symmetry in the input values $k_j$ for $1\le j \le \mathbf N,$ the knowledge
of the ratio $s_1/r_1$ in both SK1 and SK2 cases, allows us to uniquely determine all the ratios $s_j/r_j$ for $1\le j \le \mathbf N.$
By using those ratios in \eqref{5.1}, we express each $\gamma_j$ and $\gamma_j^\ast$ in terms of the real parameter $r_j$ and the
elements in the set $\{k_j,k_j^\ast\}_{j=1}^{\mathbf N}.$ 
In analogy with \eqref{5.11}, we define the coefficients $\nu_{j0},$  $\nu_{0j},$ and $\nu_{jj}$ as
\begin{equation}\label{5.30}
\nu_{j0}:=\nu\bigg|_{k_j\mapsto zk_j},\quad
\nu_{0j}:=\nu\bigg|_{k^\ast_j\mapsto z^2k^\ast_j},\quad
\nu_{jj}:=\nu\bigg|_{\substack{k_j\mapsto zk_j\\
k^\ast_j\mapsto z^2k^\ast_j}},
\end{equation}
and in analogy with \eqref{5.18} we define the coefficients $\mu_{j0},$  $\mu_{0j},$ and $\mu_{jj}$ as
\begin{equation}\label{5.31}
\mu_{j0}:=\mu\bigg|_{k_j\mapsto zk_j},\quad
\mu_{0j}:=\mu\bigg|_{k^\ast_j\mapsto z^2k^\ast_j},\quad
\mu_{jj}:=\mu\bigg|_{\substack{k_j\mapsto zk_j\\
k^\ast_j\mapsto z^2k^\ast_j}}.
\end{equation}
Similarly, in analogy with \eqref{5.23} and \eqref{5.29}, we define the quantities
$\alpha_j$ and $\beta_j$ as
\begin{equation}\label{5.32}
\alpha_j:=(\Sigma^2_{\mathbf N}-\Pi_{\mathbf N})-(k_j+k^\ast_j)\,\Sigma_{\mathbf N}+
(k_j+k^\ast_j)^2,
\end{equation}
\begin{equation}\label{5.33}
\beta_j:=(\Sigma^2_{\mathbf N}-\Pi_{\mathbf N})-(k_j+k^\ast_j)\,\Sigma_{\mathbf N}.
\end{equation}
Then, using \eqref{5.30}--\eqref{5.33}, in analogy with \eqref{5.22} and \eqref{5.28}, respectively, we obtain the ratios $s_j/r_j$ in the SK1 case as
\begin{equation}\label{5.34}
\displaystyle\frac{s_j}{r_j}=i\,\displaystyle\frac{\alpha_j\,(\nu_{j0}+\nu_{0j})+(\mu_{j0}+\mu_{0j})
}{\alpha_j\,(\nu_{j0}-\nu_{0j})+(\mu_{j0}-\mu_{0j})},\qquad 1\le j\le\mathbf N,
\end{equation}
and we obtain the ratios $s_j/r_j$ in the SK2 case as
\begin{equation}\label{5.35}
\displaystyle\frac{s_j}{r_j}=i\,\displaystyle\frac{\beta_j\,(\nu_{j0}+\nu_{0j})+(\mu_{j0}+\mu_{0j})
}{\beta_j\,(\nu_{j0}-\nu_{0j})+(\mu_{j0}-\mu_{0j})},\qquad 1\le j\le\mathbf N.
\end{equation}
We then recover $Q$ by using \eqref{5.6}, where $\det[\mathbf M(x)]$ is uniquely
determined by the input data set $\{k_j,r_j\}_{j=1}^{\mathbf N}.$ The constructed $Q$ becomes the $\mathbf N$-soliton solution to the
Sawada--Kotera equation \eqref{1.1}. We remark that although the parameters $r_j$ for $1\le j \le \mathbf N$ are each real, the range
values for each $r_j$ may need to be restricted to ensure that $Q$ does not have any singularities. We deal with
those restrictions 
by using the following simple procedure and by introducing the positive parameter
$c_j$ instead of using $r_j$ for each value of $j$ for $1\le j\le\mathbf N.$

We remark that the use of the ratios $s_j/r_j$ in both cases of SK1 and SK2 not only ensures that the resulting $\det[\mathbf M(x)]$ via
\eqref{5.6} yields the potential $Q$ satisfying \eqref{1.1}, but it  also results in a significant simplification in $\det[\mathbf M(x)].$ The simplification occurs as follows.
The ratio $\det[\mathbf M(x)]/ \nu,$ where $\nu$ is the constant appearing in \eqref{5.9} and \eqref{5.10}, becomes the square of a polynomial of degree $\mathbf N$
in $\chi_1,$ $\dots,$ $\chi_{\mathbf N},$ where each $\chi_j$ for $1\le j \le \mathbf N$ appears at most to the first
power. Furthermore, that polynomial reduces to 1 when we let $\chi_j=0$ for $1\le j \le \mathbf N.$ We use $\Delta(x)$ to denote that
polynomial. We refer the reader to \cite{ACTU2025,ACTU2026} for further details on the introduction of $\Delta(x).$

We now briefly describe the properties of $\Delta(x).$ We first remark that the justification for the existence of $\Delta(x)$ can be found in
\eqref{5.20} in the SK1 case and in \eqref{5.26} in the SK2 case. Each term on the left-hand side of \eqref{5.20} is a polynomial of degree
$2\mathbf N$ in $\chi_1,$ $\dots,$ $\chi_{\mathbf N},$ whereas the right-hand side is the ratio of a polynomial of degree $4\mathbf N$ 
to a polynomial degree $2\mathbf N.$ Hence, that right-hand side must be a polynomial of degree $2\mathbf N.$ This is assured by
choosing $\det[\mathbf M(x)]$ as
\begin{equation}\label{5.36}
\det[\mathbf M(x)]=\nu\,\Delta(x)^2.
\end{equation}
The same argument also applies for the equality in \eqref{5.26}. In other words, \eqref{5.36} must hold also in the case of SK2. We already
know each coefficient in $\det[\mathbf M(x)]$ explicitly in terms of the parameters in the data set $\{k_j,k_j^\ast,r_j\}_{j=1}^{\mathbf N}.$
Furthermore, we know the explicit expression for $\nu$ in terms of the parameters $\{k_j,k_j^\ast\}_{j=1}^{\mathbf N}.$ Thus, using
\eqref{5.36} we can determine each coefficient of $\chi_j$ in the polynomial $\Delta(x).$ For this, we proceed as follows. We would like to write
$\Delta(x)$ as
\begin{equation}\label{5.37}
\Delta(x)=1+\sum_{j=1}^{\mathbf N} c_j\,\chi_j+\cdots,
\end{equation}
where each parameter $c_j$ takes positive values and the ellipsis contains the terms with the second and higher powers in the polynomial
in $\chi_1,$ $\dots,$ $\chi_{\mathbf N}.$ With the help of \eqref{5.8}, we see that $\det[\mathbf M(x)]$ can be written as
\begin{equation}\label{5.38}
\det[\mathbf M(x)]=\nu+\sum_{j=1}^{\mathbf N} \left[\nu_{j0}\,\gamma_j+\nu_{0j}\,\gamma_j^\ast \right]\chi_j+\cdots,
\end{equation}
where the ellipsis contains the terms with the second and higher powers in the polynomial in $\chi_1,$ $\dots,$ $\chi_{\mathbf N}$ and it is
understood that $\gamma_j$ and $\gamma_j^\ast$ on the right-hand side are evaluated as in \eqref{5.1} by using the ratios $s_j/r_j$ given by
\eqref{5.34} and \eqref{5.35}, respectively, in the SK1 case and in the SK2 case. After some simplification, we evaluate the coefficient of $\chi_j$ in
\eqref{5.38} in the SK1 case as
\begin{equation}
\label{5.39}
\nu_{j0}\,\gamma_j+\nu_{0j}\,\gamma_j^\ast=\displaystyle\frac{2\,r_j\left(
\nu_{0j}\,\mu_{j0}-
\nu_{j0}\,\mu_{0j}
\right)}{\alpha_j\left(\nu_{j0}-\nu_{0j}\right)+\left(\mu_{j0}-\mu_{0j}\right)},
\qquad 1\le j \le \mathbf N,
\end{equation}
and in the SK2 case as
\begin{equation}
\label{5.40}
\nu_{j0}\,\gamma_j+\nu_{0j}\,\gamma_j^\ast=\displaystyle\frac{2\,r_j\left(
\nu_{0j}\,\mu_{j0}-
\nu_{j0}\,\mu_{0j}
\right)}{\beta_j\left(\nu_{j0}-\nu_{0j}\right)+\left(\mu_{j0}-\mu_{0j}\right)},
\qquad 1\le j \le \mathbf N.
\end{equation}
With the help of \eqref{5.36}--\eqref{5.38}, we see that the positive parameters $c_j$ are given by
\begin{equation}
\label{5.41}
c_j=\displaystyle\frac{\nu_{j0}\,\gamma_j+\nu_{0j}\,\gamma_j^\ast}{2\nu},
\qquad 1\le j \le \mathbf N,
\end{equation}
with the understanding that $\gamma_j$ and $\gamma_j^\ast$ on the right-hand side are evaluated as in \eqref{5.1} by using the ratios $s_j/r_j$
given in \eqref{5.34} and \eqref{5.35}, respectively, in the SK1 and SK2 cases. 
Using \eqref{5.39}--\eqref{5.41}, in the SK1 case, we obtain
\begin{equation}\label{5.42}
c_j=\displaystyle\frac{r_j}{\nu}\left[\displaystyle\frac{\left(\nu_{0j}\,\mu_{j0}-\nu_{j0}\,\mu_{0j}\right)}{\alpha_j\left(\nu_{j0}-\nu_{0j}\right)+\left(\mu_{j0}-\mu_{0j}\right)}\right],
\qquad 1\le j \le \mathbf N,
\end{equation}
and in the SK2 case, we get
\begin{equation}\label{5.43}
c_j=\displaystyle\frac{r_j}{\nu}\left[\displaystyle\frac{\left(\nu_{0j}\,\mu_{j0}-\nu_{j0}\,\mu_{0j}\right)}{\beta_j\left(\nu_{j0}-\nu_{0j}\right)+\left(\mu_{j0}-\mu_{0j}\right)}\right],
\qquad 1\le j \le \mathbf N.
\end{equation}
From \eqref{5.42} and \eqref{5.43}, we see that each parameter $c_j$ in \eqref{5.37} is directly proportional to $r_j.$ Hence, in expressing the
$\mathbf N$-soliton solution to the Sawada--Kotera equation \eqref{1.1}, we can use the input $\{k_j,c_j\}_{j=1}^{\mathbf N}$ instead of using the
input $\{k_j,r_j\}_{j=1}^{\mathbf N}.$ As indicated already, the advantage of using the input $\{k_j,c_j\}_{j=1}^{\mathbf N}$ is that the quantity
$\Delta(x)$ becomes nonsingular 
for $x\in\mathbb R$ for each $t\in\mathbb R$ when we choose each $c_j$ as an arbitrary positive parameter.

Using the value of $c_j$ in \eqref{5.37}, we obtain the explicit expression for $\Delta(x)$ as
\begin{equation}\label{5.44}
\begin{split}
\Delta(x)=&1+\sum_{j=1}^{\mathbf N}
c_j\,\chi_j+\sum_{1\le j_1<j_2\le \mathbf N} A_{j_1 j_2}\,c_{j_1} c_{ j_2}\,\chi_{j_1}\,\chi_{j_2}\\
&+\sum_{1\le j_1<j_2<j_3\le \mathbf N} A_{j_1 j_2}\,A_{j_1 j_3}\,A_{j_2 j_3}\,c_{j_1}\, c_{ j_2}\,c_{j_3}\,
\chi_{j_1}\,\chi_{j_2}\,\chi_{j_3}\\
&+\cdots+
\left[A_{12}\,A_{13}\cdots \,A_{(\mathbf N-1)\mathbf N}\,c_{1}\,c_{2}\cdots \,c_{\mathbf N}\right]
\chi_1\,\chi_2\cdots\,\chi_{\mathbf N},
\end{split}
\end{equation}
where the right-hand side is a polynomial of degree $\mathbf N$ in $\chi_1,$ $\dots,$ $\chi_{\mathbf N}$ with the last term containing the product
$\chi_1\,\chi_2\cdots\chi_{\mathbf N}.$ The double-indexed quantity $A_{jl}$ is determined by the elements in the set $\{k_j\}_{j=1}^{\mathbf N}$
explicitly and is given by 
\begin{equation}\label{5.45}
A_{jl}:=\displaystyle\frac{(k_j-k_l)^3(k_j^3+k_l^3)}
{(k_j+k_l)^3(k_j^3-k_l^3)},
\qquad 1\le j<l \le\mathbf N.
\end{equation}

We recall that the $k_j$-values are chosen as distinct and as in \eqref{4.3}, where each $\eta_j$ is a positive parameter. Hence, the quantity $A_{jl}$ in \eqref{5.45}
is strictly positive. Thus, all the coefficients in \eqref{5.44} are positive and consequently $\Delta(x)$ itself is positive for all $x\in\mathbb R$ and each
fixed $t\in\mathbb R.$ We refer the reader to Section 6 of \cite{ACTU2026} for further details.

Using \eqref{5.36} in \eqref{5.6}, we see that the $\mathbf N$-soliton solution to the Sawada--Kotera equation \eqref{1.1} is obtained as
\begin{equation}
\label{5.46}
Q(x)=6\,\displaystyle\frac{d}{dx}\left(\displaystyle\frac{\Delta'(x)}
{\Delta(x)}\right).
\end{equation}
We remark that we have used the input $\{k_j,c_j\}_{j=1}^{\mathbf N}$ to construct $\Delta(x).$ We refer to the positive parameter $c_j$ appearing
in $\Delta(x)$ as the bound-state constant at the bound state at $k=k_j.$ Instead of using 
the complex-valued parameters $k_j$ for $1\le j \le \mathbf N,$ we can use the positive
parameters $\eta_j$ appearing in \eqref{4.3}. From \eqref{4.3} we see that
each $k_j$ is a constant multiple of $\eta_j.$ Consequently, we can write $A_{jl}$ in \eqref{5.45} as
\begin{equation*}
A_{jl}:=\displaystyle\frac{(\eta_j-\eta_l)^3(\eta_j^3+\eta_l^3)}
{(\eta_j+\eta_l)^3(\eta_j^3-\eta_l^3)},
\qquad 1\le j<l \le\mathbf N.
\end{equation*}
Thus, the quantity $\Delta(x)$ can also be uniquely constructed by using the input data set
$\{\eta_j,c_j\}_{j=1}^{\mathbf N},$ where each of the parameters $\eta_j$ and $c_j$ is positive for $1\le j \le \mathbf N.$

\section{The Marchenko integral equation}
\label{section6}

In this section, we present the Marchenko integral equation related to the 
$\mathbf N$-soliton solution of the Sawada--Kotera equation \eqref{1.1}.
We recall that we use the input  data set
$\{k_j,E_j\}_{j=1}^{\mathbf N}$ or one of its equivalents in \eqref{1.2}, where our input data set
is related to the scattering data of the third-order equation \eqref{2.4} in the reflectionless case
and the left transmission coefficient $T_{\text{\rm{l}}}(k)$
is given in \eqref{4.1}. We also recall that $k_j$ and $k_j^\ast$ correspond to the bound-state locations
in the complex $k$-plane and that $E_j$ and $E_j^\ast$ denote the respective bound-state dependency constants
for $1\le j\le\mathbf N.$ After establishing our Marchenko integral equation, we show
how the $\mathbf N$-soliton solution to \eqref{1.1} is obtained from the solution to
our Marchenko integral equation. As indicated in Section~\ref{section1}, the establishment of the
Marchenko integral equation is a significant mathematical result.

\begin{figure}[!ht]
     \centering
         \includegraphics[width=1.55in,height=2.5in]{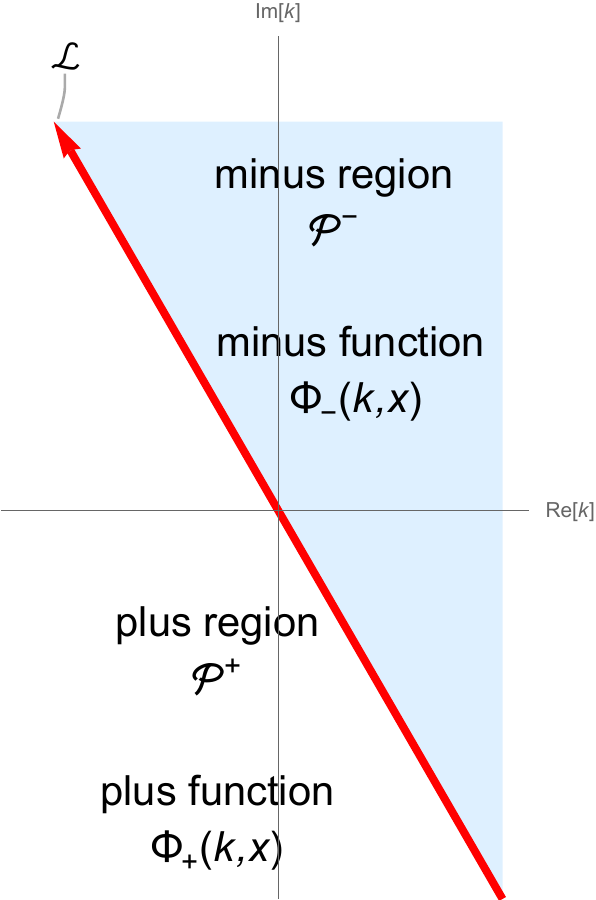}      \hskip .2in
         \includegraphics[width=3.4in,height=2.5in]{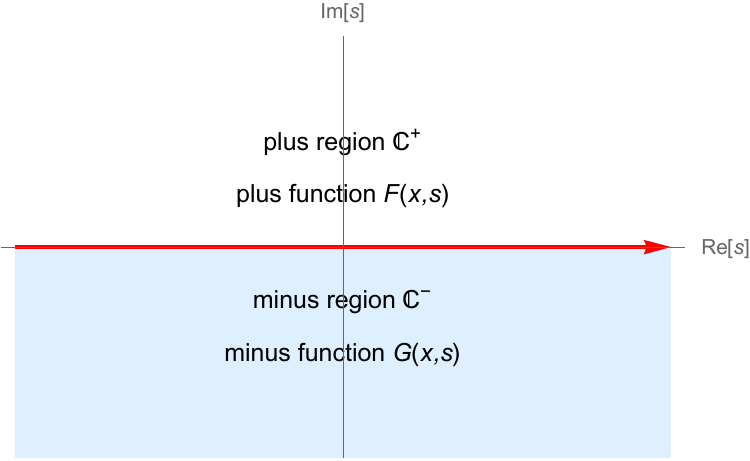} 
\caption{The mapping from the complex $k$-plane
to the complex $s$-plane transforms the plus and minus regions
$\mathcal P^+$ and $\mathcal P^-$ to
$\mathbb C^+$ and $\mathbb C^-,$ respectively.
The plus and minus functions
$\Phi_+(k,x)$ and $\Phi_-(k,x)$ are the components of
a sectionally meromorphic function, respectively,
in the complex $k$-plane.
The plus and minus functions
$F(x,s)$ and $G(x,s)$ are the components of
a sectionally meromorphic function, respectively, in the complex
$s$-plane.}
\label{Figure2}
\end{figure}

For the derivation of our Marchenko integral equation, we proceed in the standard way so that the same procedure
can be used to derive the relevant Marchenko integral equations associated with the soliton solutions to the
Kaup--Kupershmidt equation, the bad Boussinesq equation, 
the good Boussinesq equation, and
the modified bad Boussinesq equation. We start with the Riemann--Hilbert problem formulated in
\eqref{4.9}.
By multiplying both sides of \eqref{4.9} with $e^{-kx}$ and subtracting 1 from each side, we obtain 
\begin{equation}\label{6.1}
e^{-kx}\,\Phi_+(k,x)-1=e^{-kx}\,\Phi_-(k,x)-1, \qquad k\in\mathcal L.
\end{equation}
By using the transformation $k\mapsto s,$ where $k=zs$ with $z$ being the special constant defined in \eqref{3.5}, we map the complex $k$-plane onto the complex
$s$-plane in a one-to-one and onto manner.
We remark that the transformation can also be interpreted as a clockwise  rotation in the complex plane  around the origin by $2\pi/3$ radians.
We note that the plus and minus regions $\mathcal P^+$ and 
$\mathcal P^-$ shown on the left plot in Figure~\ref{Figure2} are mapped to $\mathbb C^+$ and $\mathbb C^-,$ respectively, in the
complex $s$-plane. Here, we use $\mathbb C^+$ and $\mathbb C^-$ 
for the upper-half and lower-half
complex planes and we let $\overline{\mathbb C^+}:=\mathbb C^+\cup\mathbb R$ and
$\overline{\mathbb C^-}:=\mathbb C^-\cup\mathbb R.$ The directed full line $\mathcal L$ is mapped to the
directed real axis $\mathbb R$ in the complex $s$-plane.

By using $k=zs$ and letting
\begin{equation*}
F(x,s):=e^{-zsx}\Phi_+(zs,x)-1, \qquad s\in\mathbb R,
\end{equation*}
\begin{equation}\label{6.3}
G(x,s):=e^{-zsx}\Phi_-(zs,x)-1, \qquad s\in\mathbb R,
\end{equation}
as shown in Figure~\ref{Figure2},
we transform \eqref{6.1} to the corresponding Riemann--Hilbert problem in the complex $s$-plane, and we obtain
\begin{equation}\label{6.4}
F(x,s)=G(x,s), \qquad s\in\mathbb R.
\end{equation}
From the properties of $\Phi_+(k,x)$ and $\Phi_-(k,x)$ described in Section~\ref{section4}, it follows that,
for each fixed
$x\in\mathbb R,$ the quantity $F(x,s)$ is meromorphic with simple poles at $s=z^2k_j$ and $s=z^2k_j^\ast$ 
for $1\le j\le \mathbf N$ in $s\in\mathbb C^+$ and continuous in $s\in\overline{\mathbb C^+},$ and we have
$F(x,s)=O(1/s)$ as $s\to\infty$ in $s\in\overline{\mathbb C^+}.$ Similarly, for each fixed $x\in\mathbb R,$ it follows that the quantity $G(x,s)$ is analytic in
$s\in\mathbb C^-$ and continuous in $s\in\overline{\mathbb C^-},$ and we have $G(x,s)=O(1/s)$ as $s\to\infty$ in $s\in\overline{\mathbb C^-}.$
Consequently, the Fourier transforms in $L^2(\mathbb R)$ with $s\in\mathbb R$ for $F(x,s)$ and $G(x,s)$ exist. We define the Fourier 
transforms $\hat F(x,y)$ and $\hat G(x,y),$ respectively, by letting 
\begin{equation}\label{6.5}
\hat F(x,y):=\ds\frac{1}{2\pi} \int_{-\infty}^\infty ds\, e^{-isy}\,F(x,s), \qquad y\in\mathbb R,
\end{equation}
\begin{equation}\label{6.6}
\hat G(x,y):=\ds\frac{1}{2\pi} \int_{-\infty}^\infty ds\, e^{-isy}\,G(x,s), \qquad y\in\mathbb R.
\end{equation}
From \eqref{6.5} and \eqref{6.6}, respectively, we get the inverse Fourier transforms given by
\begin{equation*}
F(x,s)=\int_{-\infty}^\infty dy\, e^{isy}\,\hat F(x,y), \qquad s\in\mathbb R,
\end{equation*}
\begin{equation}\label{6.8}
G(x,s)=\int_{-\infty}^\infty dy\, e^{isy}\,\hat G(x,y), \qquad s\in\mathbb R.
\end{equation}
From \eqref{6.4}--\eqref{6.6} it follows that we have
\begin{equation}\label{6.9}
\hat F(x,y)=\hat G(x,y), \qquad y\in\mathbb R.
\end{equation}
Using the aforementioned properties of $G(x,s)$ when $s\in\overline{\mathbb C^-},$ from \eqref{6.8} we conclude that
\begin{equation}\label{6.10}
\hat G(x,y)=0, \qquad y>0.
\end{equation}
Then, \eqref{6.9} and \eqref{6.10} yield
\begin{equation}\label{6.11}
\hat F(x,y)=0, \qquad y>0.
\end{equation}
Thus, using \eqref{6.10} in \eqref{6.8} we obtain
\begin{equation}\label{6.12}
G(x,s)=\int_{-\infty}^0 dy\, e^{isy}\,\hat G(x,y), \qquad s\in\mathbb R.
\end{equation}

Next, we would like to evaluate $\hat G(x,y)$ when $y<0.$ In order to accomplish this, it is sufficient to determine
$\hat F(x,y)$ for $y<0$ and then use \eqref{6.9}.
For the determination of $\hat F(x,y)$ for $y<0,$ we use \eqref{6.5} when $y<0$
by evaluating the integral there with the help of the residues of the integrand at $s=z^2k_j$ and $s=z^2k_j^\ast$
for $1\le j\le\mathbf N.$
In terms of the quantity $\Gamma(k)$ in \eqref{4.2}, we define the related quantity $\Gamma_j(k),$ which differ from
$\Gamma(k)$ by the product $(k+k_j)(k+k_j^\ast).$ Thus, we let
\begin{equation}\label{6.13}
\Gamma_j(k):=\ds\frac{\Gamma(k)}{(k+k_j)(k+k_j^\ast)},\qquad 1\le j\le\mathbf N.
\end{equation}
We remark that, in analogy with the first equality of \eqref{4.18}, the quantity $\Gamma_j(k)$ also satisfies
\begin{equation}\label{6.14}
\Gamma_j(k^\ast)=\Gamma_j(k)^\ast,\qquad k\in\mathbb C.
\end{equation}
From \eqref{4.1} and \eqref{6.13}, we see that the left transmission coefficient can be written in terms of
$\Gamma_j(k)$ as
\begin{equation}\label{6.15}
T_{\text{\rm{l}}}(k)=\ds\frac{\Gamma(k)}{(k-k_j)(k-k_j^\ast)\,\Gamma_j(-k)},\qquad k\in\mathbb C.
\end{equation}
Using \eqref{6.15}, we express the value of the left transmission coefficient
in the complex $s$-plane as
\begin{equation}\label{6.16}
T_{\text{\rm{l}}}(zs)=\ds\frac{z\,\Gamma(zs)}{(s-z^2k_j)(s-z^2k_j^\ast)\,\Gamma_j(-zs)},\qquad s\in\mathbb C,
\end{equation}
where we have used the fact that $z^3=1.$
The expression in \eqref{6.16} allows us to determine the residues of
$T_{\text{\rm{l}}}(zs)$ at the poles $s=z^2k_j$ and $s=z^2k_j^\ast,$ respectively, as
\begin{equation}\label{6.17}
\mathrm{Res}[T_{\text{\rm{l}}}(zs),s=z^2k_j]=\ds\frac{z^2\,\Gamma(k_j)}{(k_j-k_j^\ast)\,\Gamma_j(-k_j)},
\end{equation}
\begin{equation}\label{6.18}
\mathrm{Res}[T_{\text{\rm{l}}}(zs),s=z^2k_j^\ast]=-\ds\frac{z^2\,\Gamma(k_j^\ast)}{(k_j-k_j^\ast)\,\Gamma_j(-k_j^\ast)}.
\end{equation}
With the help of \eqref{6.17} and \eqref{6.18}, we evaluate \eqref{6.5} for $y<0,$ and we get 
\begin{equation}\label{6.19}
\begin{split}
\hat F(x,y)=i\ds\sum_{j=1}^{\mathbf N}
\mathrm{Res}&[T_{\text{\rm{l}}}(zs),s=z^2 k_j]\,e^{-k_jx}\,f(k_j,x)\,e^{-iz^2 k_jy}
\\ &+i\ds\sum_{j=1}^{\mathbf N}\mathrm{Res}[T_{\text{\rm{l}}}(zs),s=z^2 k_j^\ast]
\,e^{-k_j^\ast x}\,f(k_j^\ast ,x)\,e^{-iz^2 k_j^\ast y}.
\end{split}
\end{equation}
Using \eqref{6.17} and \eqref{6.18} in \eqref{6.19}, we express $\hat F(x,y)$ for $y<0$ as
\begin{equation}\label{6.20}
\hat F(x,y)=iz^2 \ds\sum_{j=1}^{\mathbf N} \left(
\ds\frac{\Gamma(k_j)\,e^{-k_jx}}{(k_j-k_j^\ast)\,\Gamma_j(-k_j)}\,f(k_j,x)\,e^{-iz^2 k_jy}
-\ds\frac{\Gamma(k_j^\ast)\,e^{-k_j^\ast x}}{(k_j-k_j^\ast)\,\Gamma_j(-k_j^\ast)}\,f(k_j^\ast ,x)\,e^{-iz^2 k_j^\ast y}
\right).
\end{equation}
Next, we use \eqref{6.9} on the left-hand side and use \eqref{4.6} and \eqref{4.7} on the right-hand side of
\eqref{6.20}. This yields the expression for $\hat G(x,y)$ for $y<0$ as
\begin{equation}\label{6.21}
\hat G(x,y)=iz^2 \ds\sum_{j=1}^{\mathbf N} \left(
\ds\frac{\Gamma(k_j)\,D_j\,e^{-k_jx}}{(k_j-k_j^\ast)\,\Gamma_j(-k_j)}\,g(z k_j,x)\,e^{-iz^2 k_jy}
-\ds\frac{\Gamma(k_j^\ast)\,D_j^\ast\,e^{-k_j^\ast x}}{(k_j-k_j^\ast)\,\Gamma_j(-k_j^\ast) }\,g( z^2k_j^\ast ,x)\,e^{-iz^2 k_j^\ast y}
\right),
\end{equation}
where we recall that $D_j$ and $D_j^\ast$ are the time-dependent bound-state dependency constants at
$k=k_j$ and $k=k_j^\ast.$
From \eqref{4.3}, we see that we have
\begin{equation}\label{6.22}
k_j^\ast=-zk_j,\qquad 1\le j\le \mathbf N.
\end{equation}
Using \eqref{4.8}, \eqref{4.19}, \eqref{4.20}, and \eqref{6.22}, we express the right-hand side of \eqref{6.21} in terms of the
modified bound-state dependency constants $\gamma_j$ and $\gamma_j^\ast$ and the quantities $\chi_j.$
Hence,
for $y<0$ we get $\hat G(x,y)$ as
\begin{equation}\label{6.23}
\begin{split}
\hat G(x,y)=-iz^2 \ds\sum_{j=1}^{\mathbf N}&
\ds\frac{\Gamma(-zk_j)\,\gamma_j\,\chi_j}{(k_j-k_j^\ast)\,\Gamma_j(-k_j)}\,e^{-zk_jx}\,g(z k_j,x)\,e^{-iz^2 k_jy}
\\
&+iz^2\ds\sum_{j=1}^{\mathbf N}\ds\frac{\Gamma(-z^2 k_j^\ast)\,\gamma_j^\ast\,\chi_j}{(k_j-k_j^\ast)\,\Gamma_j(-k_j^\ast) }
\,e^{-z^2k_j^\ast x}\,g( z^2k_j^\ast ,x)\,e^{-iz^2 k_j^\ast y}.
\end{split}
\end{equation}
We recall that, in order to obtain the quantity $Q$ satisfying the Sawada--Kotera equation
\eqref{1.1}, we must use the modified bound-state dependency constants
$\gamma_j$ and $\gamma^\ast$ as in \eqref{5.1} by using the ratios
$s_j/r_j$ given by \eqref{5.34} and \eqref{5.35}, in the
SK1 and SK2 cases, respectively.

Let us introduce the quantity $b_j$ as
\begin{equation}\label{6.24}
b_j:=\ds\frac{iz^2\,\Gamma(-z k_j)\,\gamma_j\,\chi_j}{(k_j-k_j^\ast)\,\Gamma_j(-k_j)},\qquad
1\le j\le\mathbf N.
\end{equation}
Using \eqref{4.2}, \eqref{4.3}, \eqref{4.19}, \eqref{4.20}, and \eqref{6.14}, we see that
the complex conjugate $b_j^\ast$ is given by
\begin{equation}\label{6.25}
b_j^\ast=\ds\frac{iz\,\Gamma(-z^2 k_j^\ast)\,\gamma_j^\ast\,\chi_j}{(k_j-k_j^\ast)\,\Gamma_j(-k_j^\ast)},\qquad
1\le j\le\mathbf N.
\end{equation}
The use of \eqref{6.24} and \eqref{6.25} in \eqref{6.23} yields
\begin{equation}\label{6.26}
\hat G(x,y)= \ds\sum_{j=1}^{\mathbf N} \left(-
b_j \,e^{-zk_jx}\,g(z k_j,x)\,e^{-iz^2 k_jy}+z\,b_j^\ast
\,e^{-z^2k_j^\ast x}\,g( z^2k_j^\ast ,x)\,e^{-iz^2 k_j^\ast y}\right).
\end{equation}
With the help of \eqref{3.32}, \eqref{6.3}, and \eqref{6.22}, we get
\begin{equation}\label{6.27}
e^{-zk_jx}\,g(zk_j,x)=1+\int_{-\infty}^0 d\zeta\, e^{ik_j \zeta}\,\hat G(x,\zeta),
\end{equation}
\begin{equation}\label{6.28}
e^{-z^2 k_j^\ast x}\,g(z^2 k_j^\ast,x)=1+\int_{-\infty}^0 d\zeta\, e^{iz k_j^\ast \zeta}\,\hat G(x,\zeta).
\end{equation}
Using \eqref{6.27} and \eqref{6.28} in \eqref{6.26}, we obtain our Marchenko integral equation as
\begin{equation}\label{6.29}
\hat G(x,y)+\Omega(0,y)+\int_{-\infty}^0 d\zeta\,\hat G(x,\zeta)\,\Omega(\zeta,y)=0,\qquad y<0,
\end{equation}
where the integral kernel $\Omega(\zeta,y)$ is defined as
\begin{equation}\label{6.30}
\Omega(\zeta,y): =\ds\sum_{j=1}^{\mathbf N} \left(b_j
\,e^{ik_j\zeta-iz^2k_j y}-z\, b_j^\ast\,e^{i z k_j^\ast \zeta-iz^2k_j^\ast y}\right).
\end{equation}
We remark that we suppress the dependence on $x$ and $t$ in the integral kernel $\Omega(\zeta,y).$ As seen from \eqref{6.24} and \eqref{6.25}, the dependence on $x$ and $t$ in the integral kernel
$\Omega(\zeta,y)$ and in the nonhomogeneous term $\Omega(0,y)$ occurs through 
the quantities $\chi_j$ defined in \eqref{4.19}.
We recall that we already suppress the $t$-dependence in the potentials $Q$ and $P,$ the Jost solutions
$f(k,x)$ and $g(k,x)$, and other quantities related to \eqref{2.4}.

The Marchenko integral equation \eqref{6.29} corresponds to the ``GLM equation'' Kaup always wanted \cite{K2002} to obtain. We emphasize
 its simplicity and its resemblance to the Marchenko integral equation for other linear operators.
 The resemblance can be observed
 in the case of the Schr\"odinger equation on the full line by comparing \eqref{6.29} with (44) of \cite{AK2001}
 and also with the penultimate displayed equality on p.~123 of \cite{DT1979}
 and by comparing the nonhomogeneous term $\Omega(0,y)$ with (45) of \cite{AK2001}.

Next, we show how to recover the potential $Q$ in \eqref{2.4} from the solution $\hat G(x,y)$ to
the Marchenko integral equation \eqref{6.29}. In fact, the potential $Q$ is recovered as
\begin{equation}\label{6.31}
Q(x)=3iz\,\ds\frac{d \hat G(x,0^-)}{dx},\qquad x\in\mathbb R.
\end{equation} 
The expression in \eqref{6.31} is obtained from \eqref{3.20}, \eqref{3.29}, \eqref{6.12} with the help of the first line of \eqref{3.32}.
We refer the reader to p. 25 of \cite{ATU2025} for the details of the derivation of \eqref{6.31}.
We are interested in obtaining the solution $\hat G(x,y)$ to \eqref{6.29} when $y<0$ by using the same input data
set used to solve the Riemann--Hilbert problem \eqref{4.9}. We recall that the relevant input data set is given by
$\{k_j,E_j\}_{j=1}^{\mathbf N}$ or the
equivalent data set
$\{k_j,\gamma_j\}_{j=1}^{\mathbf N},$ where the
modified bound-state dependency constants $\gamma_j$ are chosen as in
\eqref{5.1} with the ratios $s_j/r_j$ as in \eqref{5.34} in the SK1 case and
as in \eqref{5.35} in the SK2 case.
To recover $\hat G(x,y)$ from our input data set, we proceed as follows.
From \eqref{6.30} we see that the nonhomogeneous term and the integral term
in \eqref{6.29} are spanned by the basis set
$\{e^{-iz^2k_j y},e^{-iz^2k_j^\ast y}\}_{j=1}^{\mathbf N}.$
Hence, the first term $\hat G(x,y)$ itself in the Marchenko equation \eqref{6.29} must be expressed
as a linear combination of the elements in the same basis set. Consequently, there must exist
the coefficients $g_j$ and $h_j$ for $1\le j\le\mathbf N$ satisfying
\begin{equation}\label{6.32}
\hat G(x,y)=\sum_{j=1}^{\mathbf N} \left(g_j\,e^{-iz^2k_j y}+h_j\,e^{-iz^2k_j^\ast y}\right),\qquad y<0.
\end{equation}
We remark that $g_j$ and $h_j$ are functions of $x$ and $t,$ but we suppress those dependencies in our notation.
We use \eqref{6.30} and \eqref{6.32} in \eqref{6.29}, and for $y<0$ we obtain
\begin{equation*}
\begin{split}
\sum_{j=1}^{\mathbf N} &\left(g_j\,e^{-iz^2k_j y}+h_j\,e^{-iz^2k_j^\ast y}\right)+\sum_{j=1}^{\mathbf N}  \left(b_j
\,e^{-iz^2k_j y}-z\, b_j^\ast\,e^{-iz^2k_j^\ast y}\right)\\
&+\ds\int_{-\infty}^0 d\zeta\, \sum_{l=1}^{\mathbf N} \left(g_l\,e^{-iz^2k_l \zeta}+h_l\,e^{-iz^2k_l^\ast \zeta}\right)\sum_{j=1}^{\mathbf N} \left(b_j
\,e^{ik_j\zeta-iz^2k_j y}-z\, b_j^\ast\,e^{i z k_j^\ast \zeta-iz^2k_j^\ast y}\right)=0.
\end{split}
\end{equation*}
Since the functions in the basis set
$\{e^{-iz^2k_j y},e^{-iz^2k_j^\ast y}\}_{j=1}^{\mathbf N}$
are linearly independent for $y<0,$
each coefficient of the basis elements in \eqref{6.29} must vanish. This yields
the linear algebraic system of $2\mathbf N$ equations in the
$2\mathbf N$ unknowns $g_1,h_1,\dots,g_{\mathbf N},h_{\mathbf N},$
and we get
\begin{equation}\label{6.34}
\begin{cases}
g_j+b_j+b_j \ds\sum_{l=1}^{\mathbf N}\ds\int_{-\infty}^0 d\zeta\, \left(g_l\,e^{ik_j\zeta-iz^2k_l \zeta}+h_l\,e^{i k_j \zeta-iz^2k_l^\ast \zeta}\right)=0, \qquad 1\le j\le \mathbf N,\\
\noalign{\medskip}
h_j-z\,b_j^\ast-z\,b_j^\ast \ds\sum_{l=1}^{\mathbf N}\ds\int_{-\infty}^0 d\zeta\left(g_l\,e^{i z k_j^\ast \zeta-iz^2k_l \zeta}+h_l\,e^{i z k_j^\ast \zeta-iz^2k_l^\ast \zeta}\right)=0,\qquad 1\le j\le \mathbf N.
\end{cases} 
\end{equation}
The integrals in \eqref{6.34} can be explicitly evaluated, and \eqref{6.34} becomes equivalent to
\begin{equation}\label{6.35}
\begin{cases}
g_j+\ds\sum_{l=1}^{\mathbf N} \left(\ds\frac{b_j}{ik_j-iz^2 k_l}\,g_l+\ds\frac{b_j}{ik_j-iz^2 k_l^\ast}\,h_l\right)=-b_j,\qquad 1\le j\le \mathbf N,\\
\noalign{\medskip}
h_j+ \ds\sum_{l=1}^{\mathbf N} \left(\ds\frac{-z\,b_j^\ast}{izk_j^\ast-iz^2k_l}\,g_l+
 \ds\frac{-z\,b_j^\ast}{izk_j^\ast-iz^2 k_l^\ast}\,h_l\right)=z\,b_j^\ast,\qquad 1\le j\le \mathbf N.
\end{cases} 
\end{equation}
With the help of \eqref{4.3}, we get
\begin{equation}\label{6.36}
\begin{cases}
ik_j-iz^2 k_l=-iz^2(k_j^\ast+k_l),
\\
\noalign{\medskip}
ik_j-iz^2 k_l^\ast=i(k_j+k_l),\\
\noalign{\medskip}
izk_j^\ast-iz^2 k_l=-iz^2(k_j+k_l),\\
\noalign{\medskip}
izk_j^\ast-iz^2 k_l^\ast=-iz^2(k_j+k_l^\ast).
\end{cases} 
\end{equation}
Using \eqref{6.36} in \eqref{6.35}, we write \eqref{6.35} in the equivalent form as
\begin{equation}\label{6.37}
\begin{cases}
g_j+\ds\sum_{l=1}^{\mathbf N} \left(\ds\frac{i z b_j}{k_j^\ast+k_l}\,g_l+\ds\frac{-i b_j}{k_j+k_l}\,h_l\right)=-b_j,\qquad 1\le j\le \mathbf N,\\
\noalign{\medskip}
h_j+ \ds\sum_{l=1}^{\mathbf N} \left(\ds\frac{-i z^2 \,b_j^\ast}{k_j+k_l}\,g_l+
 \ds\frac{-i z^2\,b_j^\ast}{k_j+k_l^\ast}\,h_l\right)=z\,b_j^\ast,\qquad 1\le j\le \mathbf N.
\end{cases} 
\end{equation}

We can write the linear algebraic system \eqref{6.35} in the matrix notation as
\begin{equation}\label{6.38}
\mathbf m(x)\, \mathbf a(x)=-\mathbf b(x),
\end{equation}
where $\mathbf m(x)$ is the $2\mathbf N\times 2\mathbf N$ coefficient matrix given by
\begin{equation}\label{6.39}
\mathbf m(x)
:=\begin{bmatrix}
1+\frac{i z b_1}{k_1^\ast+k_1} & \frac{-i b_1}{k_1+k_1} & \frac{i z b_1}{k_1^\ast+k_2} & \frac{-i b_1}{k_1+k_2} & \cdots & \frac{i z b_1}{k_1^\ast+k_{\mathbf N}} & \frac{-i b_1}{k_1+k_{\mathbf N}}\\
\noalign{\medskip}
\frac{-i z^2 b_1^\ast}{k_1+ k_1} & 1+\frac{-iz^2 b_1^\ast}{k_1+k_1^\ast}&\frac{-iz^2 b_1^\ast}{k_1+ k_2} & \frac{-iz^2  b_1^\ast}{k_1+ k_2^\ast} & \cdots & \frac{-i z^2 b_1^\ast}{k_1+k_{\mathbf N}} &\frac{-iz^2  b_1^\ast}{k_1+ k_{\mathbf N}^\ast}\\
\noalign{\medskip}
\frac{i z b_2}{k_2^\ast+k_1} &\frac{-i b_2}{k_2+k_1} & 1+\frac{i z b_2}{k_2^\ast+k_2} & \frac{-i b_2}{k_2+k_2} & \cdots & \frac{i z b_2}{k_2^\ast+k_{\mathbf N}} & \frac{-i b_2}{k_2+k_{\mathbf N}}\\
\noalign{\medskip}
\frac{-i z^2 b_2^\ast}{k_2+k_1} &\frac{-i z^2 b_2^\ast}{k_2+k_1^\ast} &\frac{-i z^2 b_2^\ast}{k_2+k_2} &1+\frac{-i z^2 b_2^\ast}{k_2+ k_2^\ast} & \cdots & \frac{-i z^2 b_2^\ast}{k_2+k_{\mathbf N}} &\frac{-i z^2 b_2^\ast}{k_2+k_{\mathbf N}^\ast}\\
\noalign{\medskip}
\vdots&\vdots&\vdots&\vdots&\ddots&\vdots&\vdots\\
\noalign{\medskip}
\frac{i z b_{\mathbf N}}{k_{\mathbf N}^\ast+k_1} &\frac{-i b_{\mathbf N}}{k_{\mathbf N}+k_1} & \frac{i z b_{\mathbf N}}{k_{\mathbf N}^\ast+k_2}&\frac{-i b_{\mathbf N}}{k_{\mathbf N}+k_2} & \cdots & 1+\frac{i z b_{\mathbf N}}{k_{\mathbf N}^\ast+k_{\mathbf N}} & \frac{-i b_{\mathbf N}}{k_{\mathbf N}
+k_{\mathbf N}}\\
\noalign{\medskip}
\frac{-i z^2  b_{\mathbf N}^\ast}{k_{\mathbf N}+ k_1}&\frac{-i z^2 b_{\mathbf N}^\ast}{k_{\mathbf N}+k_1^\ast} & \frac{-i z^2 b_{\mathbf N}^\ast}{k_{\mathbf N}+k_2}& \frac{-i z^2 b_{\mathbf N}^\ast}{k_{\mathbf N}+k_2^\ast} & \cdots & \frac{-i z^2  b_{\mathbf N}^\ast}{k_{\mathbf N}+ k_{\mathbf N}} &1+ \frac{-i z^2 b_{\mathbf N}^\ast}{k_{\mathbf N}+k_{\mathbf N}^\ast}
\end{bmatrix},
\end{equation}
and $\mathbf a(x)$ and $\mathbf b(x)$ are the column vectors with $2\mathbf N$ entries, and they are given by
\begin{equation}\label{6.40}
\mathbf a(x): =\begin{bmatrix}
g_1\\
h_1\\
g_2\\
h_2\\
\vdots\\
g_{\mathbf N}\\
h_{\mathbf N}\\
\end{bmatrix},
\quad
\mathbf b(x):=\begin{bmatrix}
b_1\\
-z b_1^\ast\\
b_2\\
-z b_2^\ast\\
\vdots\\
b_{\mathbf N}\\
-z b_{\mathbf N}^\ast\\
\end{bmatrix}.
\end{equation}

From \eqref{6.38} we recover the unknown column vector $\mathbf a(x)$ as
\begin{equation}\label{6.41}
\mathbf a(x)=-\mathbf m(x)^{-1}\,\mathbf b(x).
\end{equation}
Having obtained $\mathbf a(x),$ we use in \eqref{6.32} its components $g_j$ and $h_j$ for $1\le j\le \mathbf N,$
and we get $\hat G(x,y).$ 
For this, we proceed as follows. We write \eqref{6.32} in the matrix form as
\begin{equation}\label{6.42}
\hat G(x,y)=\begin{bmatrix}
e^{-iz^2k_1 y}&e^{-iz^2k_1^\ast y}&\cdots&
e^{-iz^2k_{\mathbf N} y}&e^{-iz^2k_{\mathbf N}^\ast y}
\end{bmatrix}\begin{bmatrix}
g_1\\
h_1\\
g_2\\
h_2\\
\vdots\\
g_{\mathbf N}\\
h_{\mathbf N}\\
\end{bmatrix}.
\end{equation}
Using \eqref{6.40} and \eqref{6.41} in \eqref{6.42}, we obtain
\begin{equation}\label{6.43}
\hat G(x,y)=-\begin{bmatrix}
e^{-iz^2k_1 y}&e^{-iz^2k_1^\ast y}&\cdots&
e^{-iz^2k_{\mathbf N} y}&e^{-iz^2k_{\mathbf N}^\ast y}
\end{bmatrix}
\mathbf m(x)^{-1}\,\mathbf b(x).
\end{equation}
Using (15) on p.~12 of \cite{CH1989}, we can write the right-hand side of \eqref{6.43} as the ratio of two determinants, and we get
\begin{equation}\label{6.44}
\hat G(x,y)=\ds\frac{\det\begin{bmatrix}\begin{array}{c | c c c c c} 
	0 & e^{-iz^2k_1 y}&e^{-iz^2k_1^\ast y}&\cdots&
e^{-iz^2k_{\mathbf N} y}&e^{-iz^2k_{\mathbf N}^\ast y}\\ 
	\hline 
	\mathbf b(x)& && \mathbf m(x)& &
\end{array}
\end{bmatrix}}{\det[\mathbf m(x)]}.
\end{equation}
From \eqref{6.44}, we have
\begin{equation}\label{6.45}
\hat G(x,0^-)=\ds\frac{\det\begin{bmatrix}\begin{array}{c | c c c c c} 
	0 & 1&1&\cdots&
1&1\\ 
	\hline 
	\mathbf b(x)& && \mathbf m(x)& &
\end{array}
\end{bmatrix}}{\det[\mathbf m(x)]}.
\end{equation}
In fact, it can be shown that the determinant of the matrix appearing in the numerator on the right-hand side in \eqref{6.45}
is related 
to the $x$-derivative of the determinant of $\mathbf m(x).$ We have
\begin{equation}\label{6.46}
\det\begin{bmatrix}\begin{array}{c | c c c c c} 
	0 & 1&1&\cdots&
1&1\\ 
	\hline 
	\mathbf b(x)& && \mathbf m(x)& &
\end{array}
\end{bmatrix}=-i z^2\,\ds\frac{d \det[\mathbf m(x)]}{dx}.
\end{equation}
The proof of \eqref{6.46} can be given as follows. Using \eqref{6.39} and the second equality of \eqref{6.40}, we see that the determinant on the
left-hand side of \eqref{6.46} is a polynomial of degree $2\mathbf N$ in the variables $b_1,$
$b_1^\ast,$ $\dots,$ $b_{\mathbf N},$ $b_{\mathbf N}^\ast,$ where each of those
$2\mathbf N$ variables appears in the polynomial at most to the first power. From \eqref{6.39} we know that $\det[\mathbf m(x)]$ is also a polynomial of degree
$2\mathbf N$ in the variables $b_1,$
$b_1^\ast,$ $\dots,$ $b_{\mathbf N},$ $b_{\mathbf N}^\ast,$ where each of those $2\mathbf N$ variables appears in the polynomial at most
to the first power. Furthermore, from \eqref{6.24} and \eqref{6.25} we observe that each of $b_j$ and $b_j^\ast$ is a function of $x$ only through the
$x$-dependence in the term $\chi_j$ defined in \eqref{4.19} and that each of $b_j$ and $b_j^\ast$ is a constant multiple of $\chi_j.$ Hence, the $x$-derivative of
$b_j$ is a constant multiple of $b_j$ and the $x$-derivative of $b_j^\ast$ is a constant
multiple of $b_j^\ast.$ Thus, the right-hand side in \eqref{6.46} is also a polynomial of degree 
$2\mathbf N$ in the variables $b_1,$
$b_1^\ast,$ $\dots,$ $b_{\mathbf N},$ $b_{\mathbf N}^\ast,$ where each of those $2\mathbf N$ variables appears in the polynomial at most 
to the first power. In the polynomial equality in \eqref{6.46}, we observe the equivalence of the corresponding coefficients. Hence, 
we have shown that \eqref{6.46} 
holds. 

Using \eqref{6.46} in \eqref{6.45} we obtain
\begin{equation}\label{6.47}
\hat G(x,0^-)=\ds\frac{-i z^2}{\det[\mathbf m(x)]}\,\ds\frac{d \det[\mathbf m(x)]}{dx}.
\end{equation}
Finally, using \eqref{6.47} in \eqref{6.31}, we recover the potential $Q$ in \eqref{2.4} as
\begin{equation}\label{6.48}
Q(x)=3\,\ds\frac{d}{dx}\left[\ds\frac{1}{\det[\mathbf m(x)]}\,\ds\frac{d \det[\mathbf m(x)]}{dx}\right].
\end{equation}

It can be verified that the determinant of the matrix $\mathbf m(x)$ appearing in \eqref{6.39} is related to the determinant
of the matrix $\mathbf M(x)$ appearing in \eqref{4.23}, and we have
\begin{equation}\label{6.49}
\det[\mathbf M(x)]=\nu\,\det[\mathbf m(x)],
\end{equation}
where $\nu$ is the Vandermonde coefficient appearing in \eqref{5.9} and \eqref{5.10}.
The proof of \eqref{6.49} is similar to the proof of \eqref{6.46}, and it can be given as follows. From \eqref{5.14}
we see that the left-hand side of \eqref{6.49} is a polynomial of degree $2\mathbf N$ in the variables
$\gamma_1\chi_1,$
$\gamma_1^\ast\chi_1,$
$\dots,$
$\gamma_\mathbf N\chi_\mathbf N,$
$\gamma_\mathbf N^\ast\chi_\mathbf N$ and that each of those $2\mathbf N$
variables appears in the polynomial at most to the first power. Moreover, from \eqref{6.24} and \eqref{6.25} we see that $\gamma_j\chi_j$ is a constant multiple
of $b_j$ and that $\gamma_j^\ast\chi_j$ is a constant multiple of $b_j^\ast.$ Hence, the left-hand side of \eqref{6.49} is a polynomial of degree $2\mathbf N$ in the
variables $b_1,$
$b_1^\ast,$
$\dots,$
$b_\mathbf N,$
$b_{\mathbf N}^\ast,$ where each of those $2\mathbf N$ variables appears at most to the first power. 
From \eqref{6.39} we see that $\det[\mathbf m(x)]$
is also a polynomial of degree $2\mathbf N$ in the variables $b_1,$
$b_1^\ast,$ $\dots,$ $b_\mathbf N,$
$b_{\mathbf N}^\ast,$
where each of those $2\mathbf N$ variables appears in the polynomial at most to the first power. 
In the polynomial equality in \eqref{6.49}, we observe the
equivalence of the corresponding coefficients. Thus, we have shown that \eqref{6.49} holds.

With the help of \eqref{6.49}, we see that \eqref{6.48} is equivalent to \eqref{5.6}.
We remark that \eqref{5.6} and \eqref{6.48} hold without restricting the ratios $s_j/r_j$ as in \eqref{5.34} or \eqref{5.35}.
Without using those restrictions, the expression $Q$ is not yet related to $P$ as in the SK1 or SK2 cases.
Only after we use those restrictions on the ratios $s_j/r_j,$ the resulting expression for $Q$ in \eqref{6.48}
satisfies the Sawada--Kotera equation \eqref{1.1} with $P\equiv 0$ in the SK1 case and with
$P=Q_x$ in the SK2 case. Thus, the equality \eqref{5.36} holds only after we impose the ratios $s_j/r_j$ in \eqref{5.34}
for the SK1 case or the ratios $s_j/r_j$ in \eqref{5.35}
for the SK2 case. Hence, after those ratios are used we get
\begin{equation}
\label{6.50}
\det[\mathbf m(x)]=\Delta(x)^2,
\end{equation}
where $\Delta(x)$ is the quantity in \eqref{5.36} and \eqref{5.44}.

Let us finally remark that in \cite{ATU2025} we have presented the Marchenko integral equation for \eqref{2.4}
 in the absence of bound states.
The recovery of the potential $Q$ from the solution to the Marchenko integral equation is given in \eqref{6.31} and that
formula holds both in the presence or absence of bound states. As indicated in the last displayed equation
in Section~5 of \cite{ATU2025}, the recovery of the potential $P$ from the solution to the Marchenko equation
is obtained via
\begin{equation*}
P(x)=-3z^2\,\hat G(x,0^-)\,\ds\frac{d\hat G(x,0^-)}{dx}+3iz\,\ds\frac{d^2\hat G(x,0^-)}{dx^2}-3z^2\,\ds\frac{d\hat G_y(x,0^-)}{dx}, \qquad x\in\mathbb R,
\end{equation*}
where $\hat G_y(x,0^-)$ denotes the value of the partial $y$-derivative of
$\hat G(x,y)$ evaluated in the limit $y\to 0^-.$

\section{Conclusion}
\label{section7}

In this paper we have presented the Marchenko integral equation \eqref{6.29} for the third-order linear equation \eqref{2.4} when the reflection coefficients are all zero. In the reflectionless case the input data set to solve the corresponding inverse scattering problem consists
of the bound-state information, which comprises the bound-state poles of a transmission coefficient and the corresponding
bound-state dependency constants. We have demonstrated how the relevant input data set yields the solution of the corresponding Riemann--Hilbert problem \eqref{4.9}. We have also shown how the relevant Marchenko integral equation is obtained by using the appropriate Fourier
transformation on the Riemann--Hilbert problem and how the solution
to the Marchenko integral equation yields the potentials in \eqref{2.4}. Furthermore, 
we have illustrated that, in the reflectionless case, the kernel of
the Marchenko integral equation becomes separable, which results in a closed-form solution to the Marchenko equation 
by using the methods from linear algebra.

The importance of the Marchenko integral equation \eqref{6.29} is that,
besides offering a solution to the relevant inverse scattering problem for
the third-order linear equation \eqref{2.4}, it yields soliton solutions to the integrable Sawada--Kotera equation \eqref{1.1}.
The method presented here explains the physical origins of the $2\mathbf N$ parameters appearing
in the $\mathbf N$-soliton solution to the Sawada--Kotera equation by showing that those parameters
come from the bound-state information for the related third-order linear problem \eqref{2.4}.
In the consideration of the Sawada--Kotera equation by using the inverse scattering transform method associated with the
third-order equation, there are two distinct cases with $P\equiv 0$ and $P=Q_x.$ In each of those two cases, we 
have indicated
how the bound-state dependency constants must be chosen so that the constructed potential $Q$ in the third-order
linear problem yields a solution to the Sawada--Kotera equation.

We have derived our Marchenko integral equation yielding the $\mathbf N$-soliton solution to the Sawada--Kotera equation \eqref{1.1} by employing 
a procedure used to derive the already known Marchenko integral equations for the second-order linear equation, i.e. for the
Schr\"odinger equation, and for various first-order linear systems such as the AKNS system \cite{AKNS1974}. This standard procedure
involves the formulation of a corresponding Riemann--Hilbert problem and then the derivation of
the Marchenko integral equation by using a Fourier transformation. Traditionally, in the formulation of a Riemann--Hilbert problem, the jump
for the sectionally analytic function is determined by both the reflection coefficients and the bound-state information.
Our own formulation of the Riemann--Hilbert problem comprises the determination of a sectionally meromorphic function
with the jump condition that is solely determined by the reflection coefficients. This latter formulation of a Riemann--Hilbert problem
is better suited to solve the inverse scattering problem
for the third-order linear operator in the reflectionless case. This fact also plays a key role in the formulation of
our Marchenko integral equation.

The technique we have developed in our paper to obtain the $\mathbf N$-soliton
solution to the Sawada--Kotera equation can readily be applied to other integrable evolution equations associated
with the third-order linear problem \eqref{2.4}. In the near future we plan to present the relevant formulation of
the Riemann--Hilbert problem and the Marchenko integral equation for the $\mathbf N$-soliton solution,
for any arbitrary positive integer $\mathbf N,$ for the 
Kaup--Kupershmidt equation
\begin{equation*}
Q_t+Q_{xxxxx}+\ds\frac{25}{2}\,Q_x Q_{xx}+5\, Q\,Q_{xxx}+ 5\,Q^2\,Q_x=0,\qquad x,t\in\mathbb R,
\end{equation*}
the bad Boussinesq equation
 \begin{equation*}
Q_{tt}-Q_{xx}+(Q^2)_{xx}-Q_{xxxx}=0,\qquad x,t\in\mathbb R,
\end{equation*}
the good Boussinesq equation
 \begin{equation*}
Q_{tt}-Q_{xx}+(Q^2)_{xx}+Q_{xxxx}=0,\qquad x,t\in\mathbb R,
\end{equation*}
and the modified bad Boussinesq equation
 \begin{equation*}
Q_{tt}+(Q^2)_{xx}-Q_{xxxx}=0,\qquad x,t\in\mathbb R,
\end{equation*}
where the $\mathbf N$-soliton solution $Q$ in each of these four cases vanishes as $x\to\pm\infty$ for each fixed $t\in\mathbb R$ and that
the $2\mathbf N$ parameters in each soliton solution is related to the bound-state
poles of a transmission coefficient and the bound-state dependency constants associated with
\eqref{2.4} with the appropriate choice of the accompanying potential $P.$

\section*{Conflicts of Interest} The authors have no conflicts of interest to declare.

\end{document}